\documentclass[namedreferences]{SolarPhysics}
\usepackage[optionalrh]{spr-sola-addons} 
\usepackage[english]{babel}
\usepackage[T1]{fontenc}
\usepackage[latin1]{inputenc}   
\usepackage{graphicx}
\usepackage{url}

\input alphabetSP 
\def\diag{{\mathrm{diag}}}

\newsavebox{\fminibox}
\newlength{\fminilength}

  \def\+{^\dagger}
\def\I{\,|\,}           

\def\nequiv{\not\kern-.05em\equiv}
\def\egal{\kern-.5em=\kern-.5em}        
\def\propt{\kern-.2em\propto\kern-.2em} 

\def\argmax{\mathop{\mathrm{arg\,max}}} 
\def\argmin{\mathop{\mathrm{arg\,min}}} 
\def\intdouble{\int\kern-0.3em\int}
\def\inttriple{\int\kern-0.3em\int\kern-0.3em\int}

\def\rond#1{\overset{\kern-0.33em~_\circ}{#1}}
\def\rondit[#1]#2{\overset{\kern#1~_\circ}{#2}}

\newcommand\sss{\scriptscriptstyle}

\newcommand\MAP{^{\sss \text{MAP}}}
\newcommand\OPT{_{\sss \text{OPT}}}
\newcommand{\grad}{ {\bf \nabla } }

\newcommand{\etal}{{\it et\ \ al.\ }}

\def\IAS{Institut d'Astrophysique Spatiale\XS}

\def\LSS{Laboratoire des Signaux et Syst\`emes\XS}

\def\CLS{Collecte Localisation Satellites\XS}

  {\def\@captype{figure}}
  {}
\makeatother

\urldef{\emailNBLSS}\url{nicolas.barbey@lss.supelec.fr}
\urldef{\emailNBIAS}\url{nicolas.barbey@ias.u-psud.fr}
\urldef{\emailFA}\url{frederic.auchere@ias.u-psud.fr}
\urldef{\emailTR}\url{thomas.rodet@lss.supelec.fr}
\urldef{\emailJCV}\url{jean-claude.vial@ias.u-psud.fr}

\begin{document}

\begin{article}

\begin{opening}

\title{A Time-Evolving 3D Method Dedicated to the Reconstruction of Solar 
  Plumes and Results Using Extreme Ultra-Violet Data}

\author{Nicolas~Barbey\footnotemark[1]\footnotemark[2]\sep
  Frédéric~Auchère\footnotemark[1]\sep
  Thomas~Rodet\footnotemark[2]\sep
  Jean~-Claude~Vial\footnotemark[1]
\footnotetext[1]{\IAS, Université Paris-Sud, Orsay, France \\
  email: \emailNBIAS, \emailFA, \emailJCV}
\footnotetext[2]{\LSS, Supéléc, Gif-sur-Yvette, France\\
 email: \emailNBLSS, \emailTR}}

\runningauthor{N.~Barbey et al.}
\runningtitle{A Time-Evolving 3D Method Dedicated to the Reconstruction of
  Solar Plumes}

\date{Received: 9 May 2007 / Accepted: 24 January 2008}

\begin{abstract}
An important issue in the tomographic reconstruction of the solar
poles is the relatively rapid evolution of the polar plumes.  We
demonstrate that it is possible to take into account this temporal
evolution in the reconstruction. The difficulty of this problem comes
from the fact that we want a 4D reconstruction (three spatial dimensions plus
time) while we only have 3D data (2D images plus time).
To overcome this difficulty, we introduce a model that describes polar plumes
as stationary objects whose intensity varies homogeneously with time.
This assumption can be physically justified if one accepts the stability of the
magnetic structure.  This model leads to a bilinear inverse problem.  We
describe how to extend linear inversion methods to these kinds of
problems.  Studies of simulations show the reliability of our method.
Results for SOHO/EIT data show that we are able to estimate the
temporal evolution of polar plumes in order to improve the reconstruction of
the solar poles from only one point of view.  We expect further improvements 
from STEREO/EUVI data when the two probes will be separated by about 60$^{\circ}$.
\end{abstract}

\end{opening}

\section{Introduction}
\label{section:Introduction}

A method known as solar rotational tomography has been used to
retrieve the 3D geometry of the solar corona (\citeauthor{Frazin00}
\citeyear{Frazin00}; \citeauthor{frazin02} \citeyear{frazin02}).
This method assumes the stability of the structures
during the time necessary to acquire the data.  Since we generally
have only one point of view at our disposal, about 15 days are
required to have data for half a solar rotation at the poles.  Here, 
we focus our study on solar polar plumes.  They are bright, radial, 
coronal ray structures located at the solar poles in regions of open magnetic
field.  The study of plumes is of great interest since it may be the
key to the understanding of the acceleration of the fast component of
the solar wind \cite{Teriaca03}.  However the three-dimensional shape
of these structures is poorly known and different assumptions have
been made, \textit{e.g.}  \citeauthor{Gabriel05} \citeyear{Gabriel05};
\opencite{Llebaria02}.
The plumes are known to evolve with a characteristic time of approximately 24
hours on spatial scales typical of 
\textit{Extreme ultra-violet Imaging Telescope}
(SOHO/EIT) data (2400 km) \cite{Deforest01}. Consequently the stability
assumption made in rotational tomography fails.  Fortunately, the
\textit{Solar TErestrial RElations Observatory} (STEREO) mission consists of
two identical spacecraft STEREO$_\text{A}$ and STEREO$_\text{B}$ which
take pictures of the Sun from two different points of view.  With the
SOHO mission still operating, this results in three,simultaneous points of view.
Three viewpoints help to improve the reconstruction of the plumes, but they are
still not enough to use standard tomographic algorithms. 
The problem is underdetermined and consequently one has to add
\textit{a priori} information in order to overcome the lack of information.
This leads to challenging and innovative signal analysis problems. 
There are different ways to deal
with underdetermination depending on the kind of object to be
reconstructed.  Interestingly the field of medical imaging faces the
same kind of issues.
In cardiac reconstruction, authors make use of
the motion periodicity in association with a high redundancy of the
data \cite{Grass03,Kachelriess00}.  If one can
model the motion as an affine transformation, and if one assumes that
we know this transformation, one can obtain an analytic solution
\cite{Ritchie96,Roux04}.

In solar tomography, the proposed innovative approaches involve the
use of additional data such as magnetic-field measurements in the
photosphere \cite{Wiegelmann03} or data fusion \cite{Frazin05c}.
Attempts have been made by 
Frazin \etal (\citeyear{Frazin05b})
to treat temporal evolution using Kalman filtering.

Since polar plumes have apparently a local, rapid, and aperiodic
temporal evolution, we developed as in the previously referenced work, a model
based on the specifics of the object we intend to reconstruct
(preliminary results can be found in
\citeauthor{Barbey07}(\citeyear{Barbey07}).  Plumes have an
intensity which evolves rapidly with time, but their position can be
considered as constant. This hypothesis is confirmed by previous
studies of the plumes such as \inlinecite{Deforest01}.  The model is made up
of an invariant morphological part $(\xb)$ multiplied by a gain term
$(\thetab_t)$ that varies with time. Only one gain term is associated
with each plume in order to constrain the model.  So we assume that
the position of each plume in the scene is known.  This model is
justified if we consider polar plumes to be slowly evolving magnetic structures
in which plasma flows.

Thanks to this model we can perform time-evolving three-dimensional
tomography of the solar corona using only extreme ultra-violet images.
Furthermore, there is no complex, underlying physical model. The only
assumptions are the smoothness of the solution, the area-dependant evolution
model, and the knowledge of the plume position. These assumptions
allow us to consider a temporal variation of a few days, while assuming
only temporal smoothness would limit variations to the order of one
solar rotation (about 27 days).  To our knowledge, the estimation of
the temporal evolution has never been undertaken in tomographic
reconstruction of the solar corona.

We first explain our reconstruction method in a Bayesian framework
(Section \ref{section:method}).  We then test the validity of our
algorithm with simulated data (Section \ref{section:validation}).  An
example of a reconstruction on real SOHO/EIT data is shown in Section
\ref{section:EITdata}.  Results are discussed in Section
\ref{section:discussion}.  We conclude in Section
\ref{section:conclusion}.

\section{Method}
\label{section:method}

Tomographic reconstruction can be seen as an inverse problem, the
direct problem being the acquisition of data images knowing the emission
volume density of the object (Section \ref{subsection:DirectProblem}).
If the object is evolving during the data acquisition, the inverse problem
is highly underdetermined.  So our first step is to redefine the
direct problem thanks to a reparametrization, in order to be able to
define more constraints (Section \ref{subsection:TemporalEvolution}).
Then, we place ourselves in the Bayesian inference framework in which
data and unknowns are considered to be random variables. The solution
of the inverse problem is chosen to be the maximum \textit{a
  posteriori} (Section \ref{subsection:InverseProblem}).  This leads
to a criterion that we minimize with an alternate optimization
algorithm (Section \ref{subsection:OptimizationAlgorithm}).

\subsection{Direct Problem}
\label{subsection:DirectProblem}

\begin{figure}
  \centering
  \includegraphics[width=.9\textwidth]{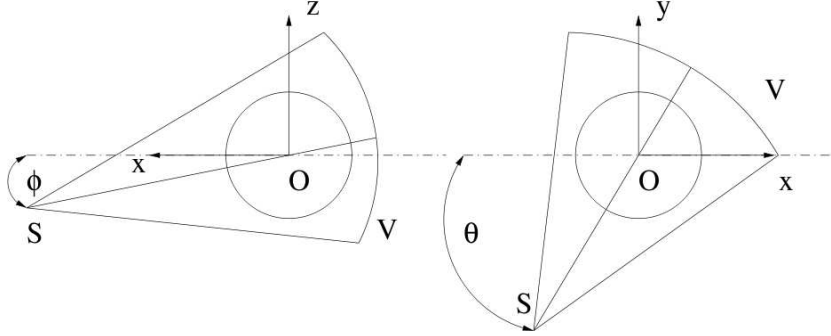}
  \caption{Scheme of the data acquisition geometry.  $(\mathbf{O};x,y,z)$
    defines the Carrington heliocentric frame of reference.  $S$ is
    the spacecraft considered. $\phi$ is the latitude, and $\theta$
    the longitude of this spacecraft.  $V$ is the virtual detector.}
  \label{fig:tomolayout}
\end{figure}

The geometrical acquisition is mathematically equivalent to a conical
beam data acquisition with a virtual spherical detector (see Figure
\ref{fig:tomolayout}).  In other words, the step between two pixels
vertically and horizontally is constant in angle.  The angle of the
full field of view is around $45$ minutes.  In order to obtain an
accurate reconstruction, we take into account the exact geometry,
which means the exact position and orientation of the spacecraft
relatively to Sun center.  We approximate integration of the emission
in a flux tube related to a pixel by an integration along the line of
sight going through the middle of that pixel.
We choose to discretize the object in the usual cubic voxels. 
$\xb$ is a vector of size $N$ containing the values of all voxels. 
In the same way, we define the
vector of data $\yb_t$ of size $M$ at time $t$.  Since the integration
operator is linear, the projection can be described by a matrix $
\Pb_t$. We choose $ \nb_t$ to be an additive noise:
\begin{equation}
  \label{eq:UsualTomographyp}
  \yb_t =  \Pb_t \xb_t +  \nb_t \text{, }\forall t \in [1, ..., T]
\end{equation}
$\Pb_t$ is the projection matrix at time $t$ of size $M \times N$
which is defined by the position and the orientation of the spacecraft
at this time.  Its transpose is the backprojection matrix. Note that
a uniform sampling in time is not required.
In order to be able to handle large problems with numerous
well-resolved data images and a large reconstruction cube, we
chose not to store the whole projection matrix. Instead, we
perform the projection operation $(\Pb\xb)$ or its transpose each time it
is needed at each iteration.  Thus, we need a very efficient
algorithm. We developed a code written in C which
performs the projection operation. It makes use of the geometrical
parameters given in the data headers in order to take into account the
exact geometry (conicity, position, and orientation of the spacecraft).
To keep this operation fast, we implemented the Siddon algorithm
\cite{Siddon85}.  It allows a fast projection or backprojection in the
case of cubic voxels (Cartesian grid). Since we focus on a small
region at the poles, we consider that we do not need to use a
spherical grid which would require a more time-consuming projection
algorithm.

We take into account the fact that the field of view is conical.
Despite the fact that the acquisition is very close to the parallel
acquisition geometry, it is sufficient to introduce an error of
several voxels of size 0.01 solar radius from one side to the other of
a three solar radii reconstructed cube.

\subsection{Modeling of the Temporal Evolution}
\label{subsection:TemporalEvolution}
With this model, the inverse problem is underdetermined since we have
at most three images at one time and we want to reconstruct the
object with its temporal evolution.  In order to do so, we first
redefine our unknowns to separate temporal evolution from spatial
structure. We introduce a new set of variables $\gb_t$ of size $N$
describing the temporal evolution and require that $\xb$ does not depend
on time:
\begin{equation}
  \label{eq:model}
  \yb_t = \Pb_t (\xb \circ \gb_t) +  \nb_t
\end{equation}
with $\circ$ being the term-by-term multiplication of vectors.  This
operator is clearly bilinear.  However, this model would increase the
number of variables excessively. 
So, we need to introduce some other kind of \textit{a priori} into our model.
We make the hypothesis that all of the voxels of one polar plume have the
same temporal evolution:
\begin{equation}
  \label{eq:localization}
  \gb_t = \Lb\thetab_t
\end{equation}
The matrix $\Lb$ of size $N \times P$ ($P$ being the number of areas)
localizes areas where the temporal evolution is identical.  Each
column of $\Lb$ is the support function of one of the plumes.  We
would like to stress that in our hypothesis, those areas do not move
relative to the object.  In other words, $\Lb$ does not depend on
time.  Localizing these areas defines $\Lb$ and only leaves $P\,T$
variables to estimate.  We redefined our problem in a way that limits
the number of parameters to estimate but still allows many solutions.
Furthermore, the problem is linear in $\xb$ knowing $\thetab$ and
linear in $\thetab$ knowing $\xb$.  It will simplify the inversion of
the problem as we shall see later. Note, however that the uniqueness of
a solution $(\xb,\thetab)$ is not guaranteed with bilinearity despite
its being guaranteed in the linear case. This example shows that $A$
can be chosen arbitrarily without changing the closeness to the data:
$\xb\circ\gb = (A\xb)\circ(A^{-1}\gb)$, where $A$ is a real
constant.  Introducing an \textit{a priori} of closeness to $\unb$ for
$\thetab$ would allow us to deal with this indeterminacy in principle.
But note that this indeterminacy is not critical since the physical
quantity of interest is only the product $\xb\circ\gb$.
\citeauthor{Feron05a} (\citeyear{Feron05a}) present a method which
solves a bilinear inversion problem in the context of microwave
tomography.

We do not deal with the estimation of the areas undergoing evolution, but we
assume in this paper that the localization is known. This localization can be
achieved using other sources of information, \textit{e.g.} stereoscopic
observations. We expect to be able to locate the areas using some other source
of information.

We can regroup the equations of the direct problem.  We have two ways
to do so, each emphasizing the linearity throughout one set of variables.
\begin{equation}
  \begin{array}{c}
    \yb = \Ub_\xb \thetab + \nb
    \\
    \left(
      \begin{array}{c}
        \yb_1\\
        \vdots\\
        \yb_T
      \end{array}
    \right)
    =
    \left(
      \begin{array}{ccc}
        \Pb_1 \Xb \Lb & & \mathbf{0}\\
        & \ddots & \\
        \mathbf{0}& & \Pb_T \Xb \Lb
      \end{array}
    \right)
    
    \left(
      \begin{array}{c}
        \thetab_1\\
        \vdots\\
        \thetab_T
      \end{array}
    \right)
    +
    \left(
      \begin{array}{c}
        \nb_1\\
        \vdots\\
        \nb_T
      \end{array}
    \right)
    
  \end{array}
\end{equation}
with $\Xb = diag(\xb)$, the diagonal matrix defined by $\xb$. $\xb$ is
of size $N$, $\yb$ and $\nb$ are of size $M\,T$, $\thetab$ is of size
$P\,T$ and $\Ub_\xb$ is of size $M\,T \times P\,T$.

Similarly,
\begin{equation}
  \begin{array}{c}
    \yb = \Vb_\thetab \xb + \nb\\
    \text{ with }
    \Vb_\thetab = 
    \left(
      \begin{array}{ccc}
        \Pb_1 \diag(\Lb\thetab_1) & & \mathbf{0}\\
        & \ddots & \\
        \mathbf{0}& & \Pb_T \diag(\Lb\thetab_T)
      \end{array}
    \right)
    \left(
      \begin{array}{c}
        \Ib_d\\
        \vdots\\
        \Ib_d
      \end{array}
    \right)
  \end{array}
\end{equation}
with $\Ib_d$ the identity matrix of size $M \times M$. $\Vb_\thetab$
is of size $MT \times N$.

\subsection{Inverse Problem}
\label{subsection:InverseProblem}

In Bayes' formalism, solving an inverse problem consists in knowing the
\textit{a posteriori} (the conditional probability density function of
the parameters, the data being given).  To do so we need to know the
likelihood (the conditional probability density function of the data
knowing the parameters) and the \textit{a priori} (the probability
density function of the parameters).  An appropriate model is a Gaussian, 
independent, identically distributed (with the same variance) noise $\nb$.
The likelihood function is deduced from the noise statistic:
\begin{equation}
  \label{eq:likelihood}
  f(\yb|\xb,\thetab,\sigma_n,\mathcal{M}) = K_1 \exp \left( - \frac{\|\yb - \Ub_\xb\thetab \|^2}{2 \sigma_\nb^2} \right)
\end{equation}
$\mathcal{M} = \left[\Pb,\Lb \right]$ describing our model (the
projection algorithm and parameters and the choice of the plume
position).  We assume that the solution is smooth spatially and
temporally, so we write the \textit{a priori} as follows:
\begin{equation}
  \label{eq:priors}
  f(\xb|\sigma_\xb) = 
  K_2 \exp \left( -\frac{\| \Db _r \xb \|^2}{2 \sigma_\xb^2} \right)
  \textrm{ and }
  f(\thetab|\sigma_\thetab) =
  K_3 \exp \left( -\frac{\| \Db _t \thetab \|^2}{2 \sigma_\thetab^2} \right)
\end{equation}
$\Db _r$ and $\Db _t$ are discrete differential operators in space and
time. Bayes' theorem gives us the \textit{a posteriori} law if we
assume that the model $\mathcal{M}$ is known as well as the
hyperparameters $\mathcal{H} = \left[ \sigma_\nb, \sigma_\xb,
  \sigma_\thetab \right]$:
\begin{equation}
  \label{eq:aposteriori}
  f(\xb,\thetab|\yb,\mathcal{H},\mathcal{M}) = 
  \frac{f(\yb|\xb,\thetab,\sigma_\nb,\mathcal{M})f(\xb|\sigma_\xb)f(\thetab|\sigma_\thetab)}
  {f(\yb|\mathcal{H},\mathcal{M})}
\end{equation}
We need to choose an estimator. It allows us to define a unique
solution instead of having a whole probability density function.  We
then choose to define our solution as the maximum \textit{a
  posteriori}.  which is given by:
\begin{equation}
  \label{eq:map}
  (\xb\MAP,\thetab\MAP) =
  \argmax_{\xb,\thetab} f(\yb|\xb,\thetab,\sigma_\nb,\mathcal{M})
  f(\xb|\sigma_\xb) f(\thetab|\sigma_\thetab)
\end{equation}
since $f(\yb|\mathcal{M})$ is a constant.  Equation (\ref{eq:map}) can
be rewritten as a minimization problem:
\begin{equation}
  \label{eq:criterion}
  (\xb\MAP,\thetab\MAP) = \argmin_{\xb,\thetab} J(\xb,\thetab)
\end{equation}
with:
\begin{equation}
  J(\xb,\thetab) =
  -2\sigma_n \log f(\xb,\thetab|\yb,\mathcal{M},\mathcal{H}) 
  =\|\yb - \Ub_\xb\thetab \|^2
  + \lambda\|\Db _r \xb \|^2 + \mu\|\Db _t\thetab\|^2\nonumber
\end{equation}
$ \lambda = \frac{\sigma_n^2}{\sigma_x^2}$ and $ \mu =
\frac{\sigma_n^2}{\sigma_\alpha^2}$ are user-defined hyperparameters.

The equivalence of Equations (\ref{eq:map}) and (\ref{eq:criterion})
has been proved by \inlinecite{Demoment89}.

Note that the solution does not have to be very smooth. It mostly
depends on the level of noise since noise increases the
underdetermination of the problem as it has been shown by the definition of
$\lambda$ and $\mu$.

\subsection{Criterion Minimization}
\label{subsection:OptimizationAlgorithm}

The two sets of variables $\xb$ and $\thetab$ are very different in
nature.  However, thanks to the problem's bilinearity, one can easily
estimate one set while the other is fixed. Consequently we perform an
iterative minimization of the criterion, and we alternate minimization
of $\xb$ and $\thetab$.  At each step $n$ we perform:
\begin{equation}
  \thetab^{n+1} = \argmin_\thetab J(\xb^n,\thetab) \text{ and }
  \xb^{n+1} = \argmin_\xb J(\xb,\thetab^{n+1})
\end{equation}

The two subproblems are formally identical. However, $\thetab$ is much
smaller than $\xb$. This is of the utmost practical importance since
one can directly find the solution on $\thetab$ by using the
pseudo-inverse method.  $\xb$ is too big for this method, and we have
to use an iterative scheme such as the conjugate-gradient to approximate the
minimum.
These standard methods are detailed in Appendices \ref{app:inverse} and \ref{app:gradient}.

\subsection{Descent Direction Definition and Stop Threshold}

We choose to use an approximation of the conjugate-gradient method
that is known to converge much more rapidly than the simple gradient
method \cite{Nocedal00,Polak69}.
\begin{equation}
  \begin{array}{ccc}
    \db^{p+1} &=& \db^p + b^p \grad_{\xb} \left. J \right|_{\xb=\xb^p} \\
    b^p &=& \frac{ \left< \grad_{\xb}\left.J \right|_{\xb=\xb^p},
        \grad_{\xb}\left. J \right|_{\xb=\xb^{p-1}} \right>}
    {\|\grad_{\xb}\left.  J \right|_{\xb=\xb^{p-1}} \|^2}
  \end{array}
  \label{eq:direction}
\end{equation}

Since the minimum is only approximately found, we need to define a
threshold which we consider to correspond to an appropriate
closeness to the data in order to stop the iterations. 
Since the solution is the point at which the
gradient is zero, we choose this threshold for updating $x$:
\begin{equation}
  \text{mean}_{\xb \in [\xb^{p},\xb^{p-1},\xb^{p-2}]}
  \| \grad_{\xb}J \|^2 < S_x
  \label{eq:Seuilx}
\end{equation}
For the global minimization, the gradient is not computed, so we
choose:
\begin{equation}
  \text{mean}_{[n,n-1,n-2]}
  \|(\xb_n,\thetab_n) - (\xb_{n-1},\thetab_{n-1}) \|^2 < S_G
  \label{eq:SeuilGlobal}
\end{equation}
Note that this way to stop the iteration
allows one to define how close one wants to be to the solution: if
the difference between two steps is below this threshold, it is
considered negligible.
The algorithm can be summarized as shown in Figure
\ref{algo:TimeTomography}.

\begin{figure}[hbtp]
  \begin{itemize}
  \item[\textbf{initialize :}] $\xb = 0$ and $\thetab = 1$
  \item[\textbf{while}] Equation (\ref{eq:SeuilGlobal}) is satisfied
    \begin{itemize}
    \item[\textit{\xb minimization:}]
    \item[\textbf{while}] Equation (\ref{eq:Seuilx}) is satisfied
      \begin{itemize}
      \item compute gradient at $\xb_n$ with Equation (\ref{eq:gradp})
      \item compute descent direction with Equation
        (\ref{eq:direction})
      \item compute optimum step with Equation (\ref{eq:step})
      \item update $\xb$ with Equation (\ref{eq:update})
      \end{itemize}
    \item[\textbf{endwhile}]
    \item[\textit{\thetab minimization:}] \
      \begin{itemize}
      \item compute the matrix $\Ub_{\xb^n}^T \Ub_{\xb^n}$ and the
        vector $\Ub_{\xb^n}^T \yb$
      \item inverse the matrix $\Ub_{\xb^n}^T \Ub_{\xb^n} + \mu
        \Db_r^T \Db_r$
      \item compute Equation (\ref{eq:thetamin})
      \end{itemize}
    \end{itemize}
  \item[\textbf{endwhile}]
  \end{itemize}
  \caption{Tomographic Reconstruction with Temporal Evolution
    Algorithm }
  \label{algo:TimeTomography}
\end{figure}

\section{Method Validation}
\label{section:validation}

In order to validate the principle of our method and test its limits,
we simulate an object containing some plumes with temporal evolution
and try to extract it from the data.

\subsection{Simulation Generation Process}
We generate an emission cube with randomly-placed, ellipsoidal plumes
with a Gaussian shape along each axis:
\begin{equation}
  \label{PlumeProfil}
  E_p = A\exp \left(
    - \frac{1}{2} \left[ \frac{\rb.\ub_\phi)}{a} \right]^2
    - \frac{1}{2} \left[ \frac{\rb.\ub_{\phi + \frac{\pi}{2}}}{b} \right]^2
  \right)
\end{equation}
The plumes evolve randomly but smoothly by interpolating over a few
randomly generated points.  Once the object is generated, we compute a
typical set of 60 images equally spaced along 180$^{\circ}$\ using our
projector algorithm.  A Gaussian random noise is added to the
projections with a signal to noise ratio (SNR) of five. The simulation
parameters are summarized in Table \ref{tab:PlumesParameters}.

\begin{table}[h!]
    \begin{tabular}{*{7}{c}}
      \hline
      Plume & Semimajor & Semiminor & $\phi$& $x_0$ & $y_0$ & Intensity \\
      Number & Axis $a$ & Axis $b$  &       &       &       & $(A)$ \\
      \hline
      1  & 4.8 & 4.2 & 1.2 & 29 & 29 & 329 \\
      \hline
      2  & 5.6 & 3.3 & 1.1 & 23 & 33 & 430 \\
      \hline
      3  & 5.2 & 4.8 & 0.1 & 40 & 42 & 723 \\
      \hline
    \end{tabular}
    \caption{Simulation Definition: Plumes Parameters}
    \label{tab:PlumesParameters}
\end{table}

\begin{table}
    \begin{tabular}{*{4}{c}}
      \hline
      cube size & cube number & pixel & projection \\
      (solar radii) & of voxels & size (radians)  & number of pixels \\
      \hline
      $1\times1\times0.05$ & $64\times64\times4$ & 
      $5\times 10^{-5}\times5\times 10^{-5}$ &$128\times8$ \\
      \hline
    \end{tabular}
    \caption{Simulation Definition: Geometric Parameters}
    \label{tab:GeometricalParameters}
\end{table}

\begin{table}
    \begin{tabular}{*{5}{c}}
      \hline
      SNR & $\lambda$ & $\mu$ & $S_\xb$ & $S_G$\\
      \hline
      $5$ & $2\times 10^{-2}$  & $100$ & $2\times 10^{-2}$& $1\times 10^{-2}$\\
      \hline
    \end{tabular}
    \caption{Simulation Definition: Other Parameters}
    \label{tab:OtherParameters}
\end{table}

\subsection{Results Analysis}
\begin{figure}
  \centering
  \begin{tabular}{ccc}
    \includegraphics[width=.27\linewidth]{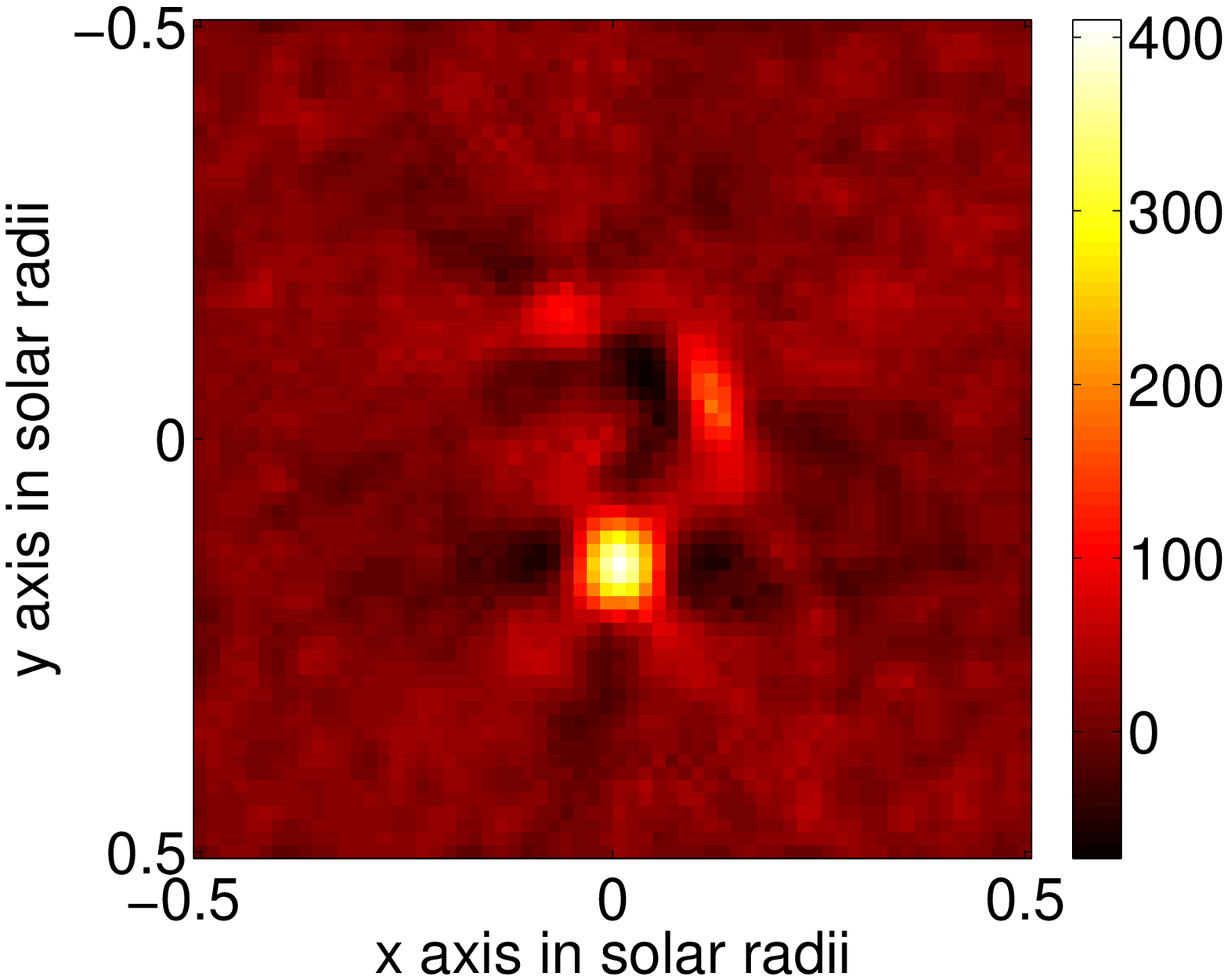}&
    \includegraphics[width=.27\linewidth]{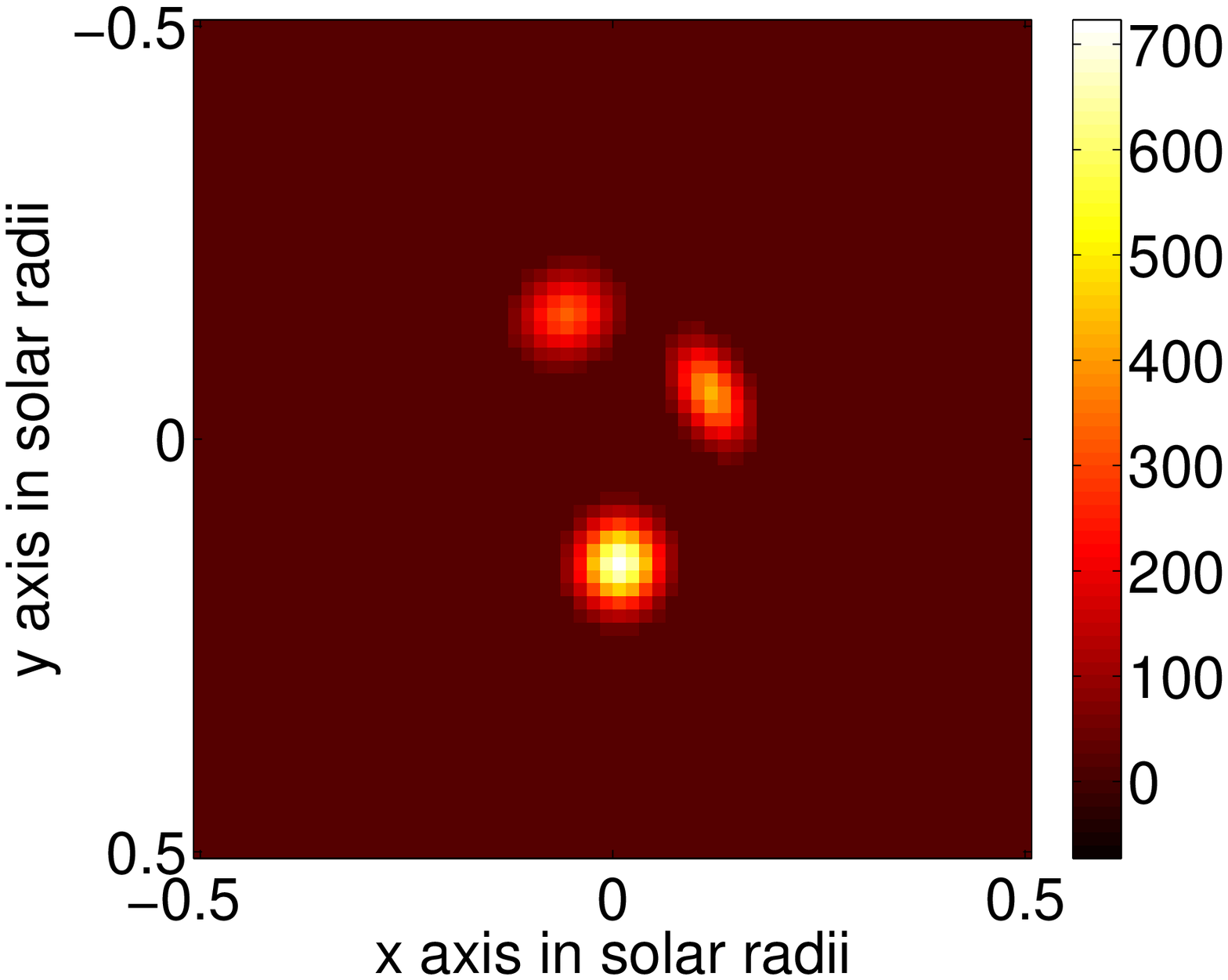}&
    \includegraphics[width=.27\linewidth]{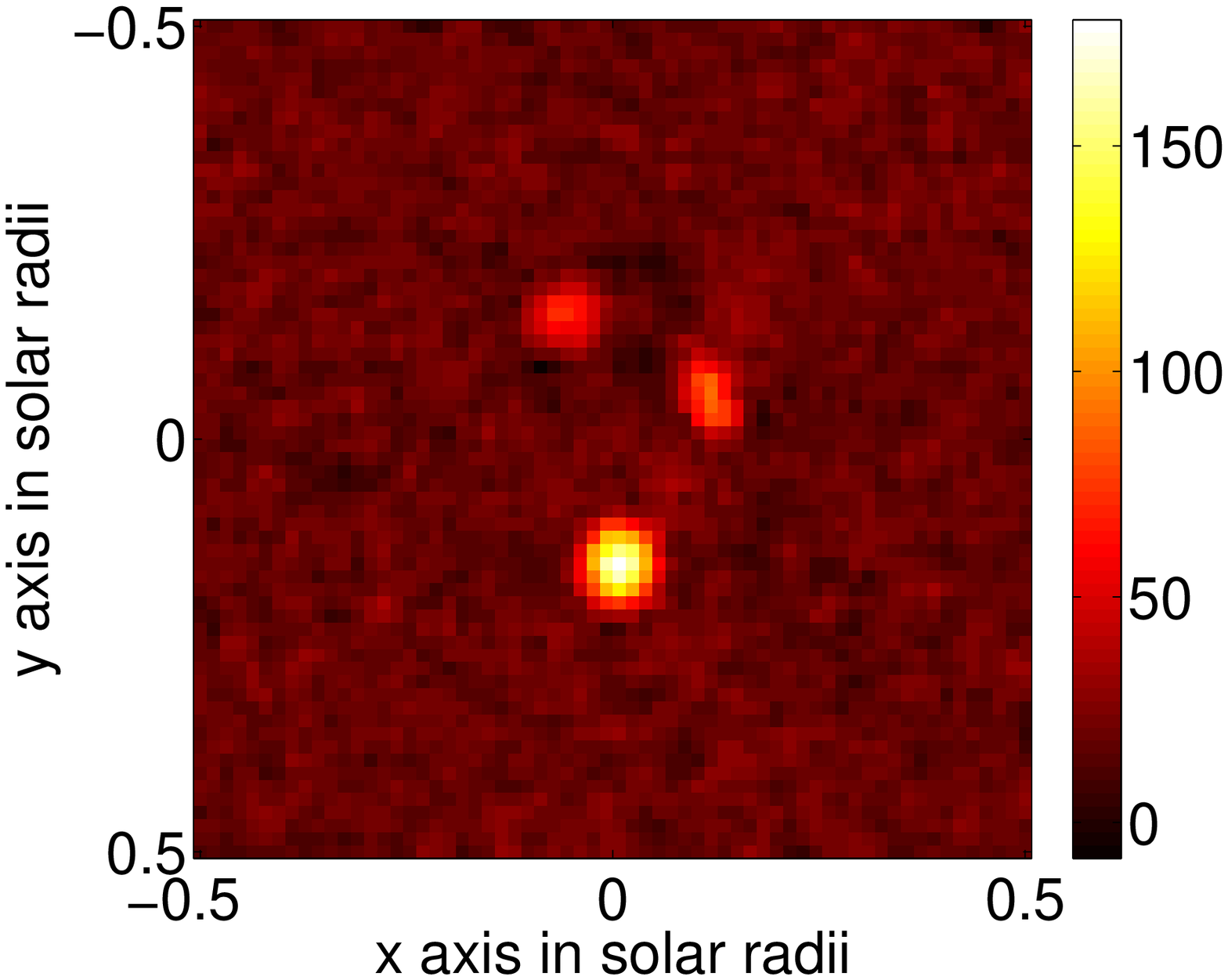}\\    
    \footnotesize{(a) $\xb$ with a FBP algorithm}&
    \footnotesize{(b) Simulated $\xb$}&
    \footnotesize{(c) $\xb$ with our algorithm}\\
    
    \includegraphics[width=.235\linewidth]{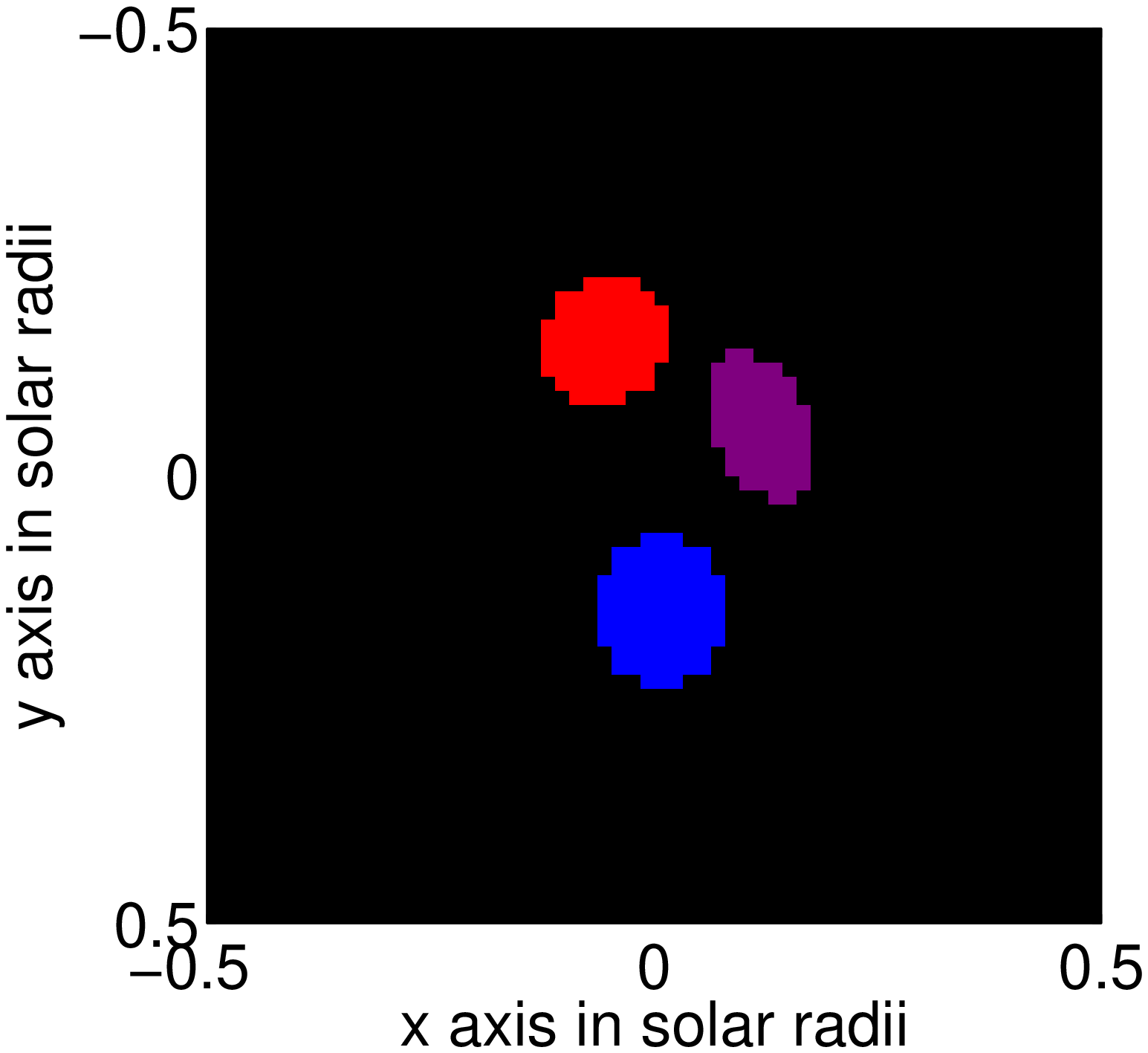}&
    \includegraphics[width=.27\linewidth]{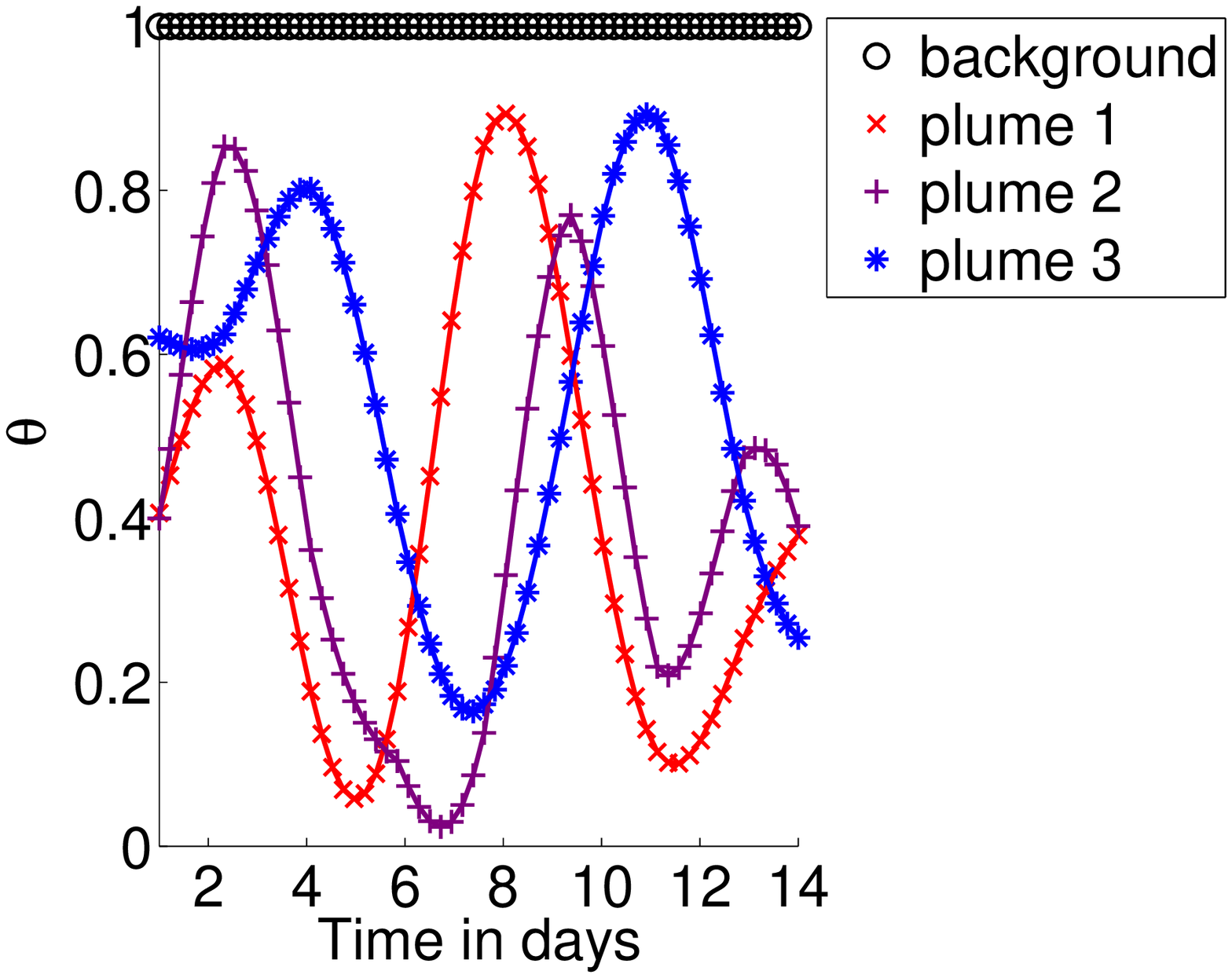}&
    \includegraphics[width=.27\linewidth]{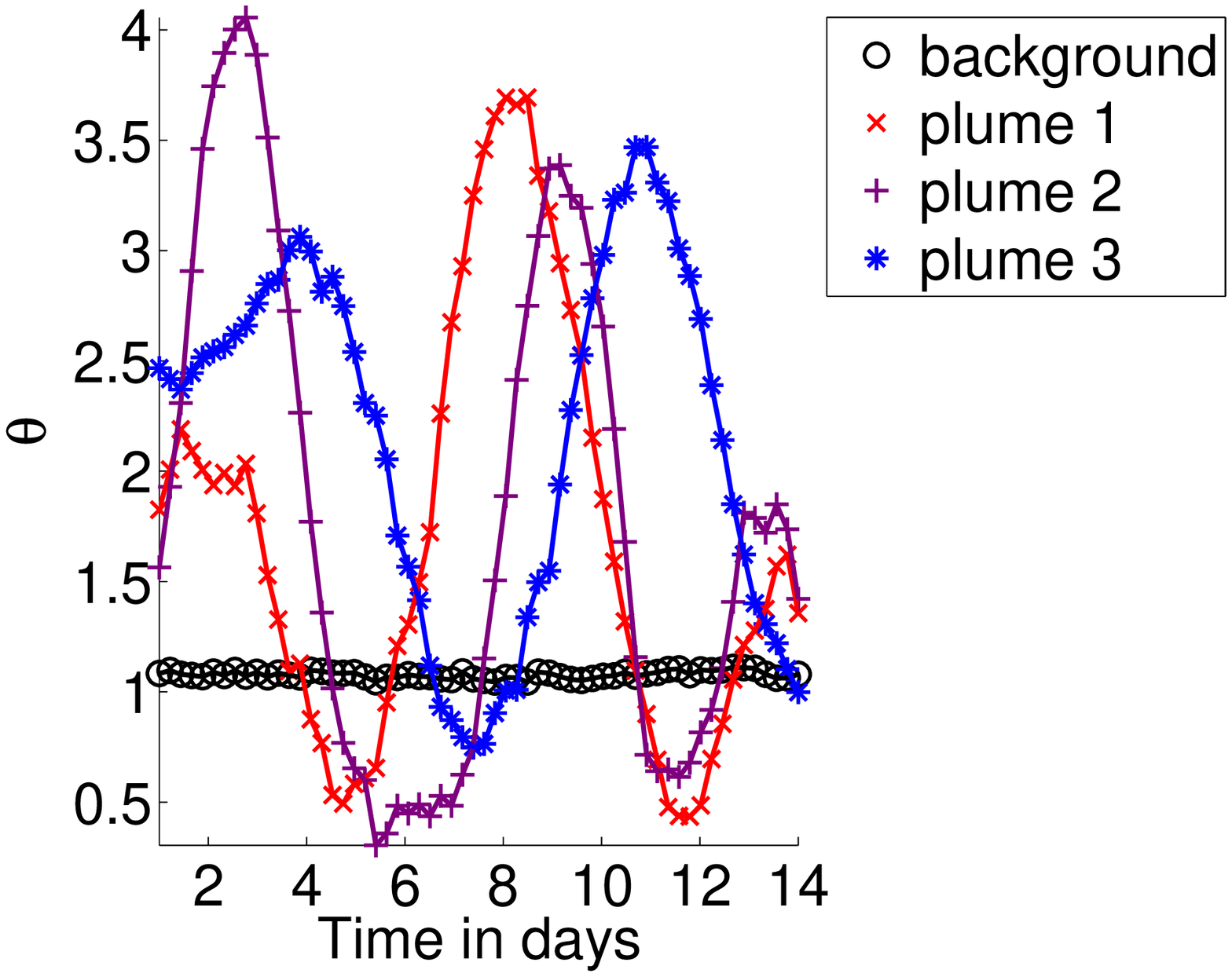}\\
    \footnotesize{(d) Chosen temporal areas}&
    \footnotesize{(e) Simulated $\thetab$}&
    \footnotesize{(f) $\thetab$ with our algorithm}
    
  \end{tabular}
  \caption
  {
    Comparison of a standard FBP method (a), the real simulated object
    (b), and the object reconstructed with our method (c).  The object
    is reconstructed using 60 projections regularly spaced over
    180$^{\circ}$.  The areas of homogeneous temporal evolution (e)
    are the same in the simulation and the reconstruction.  We
    associated one time per projection to define $\thetab$ in the
    simulation (e) and our reconstruction (f). The time scale is in
    days assuming a rotation speed of half a rotation in 14 days.
    $\xb$ is the spatial distribution of the emission density volume.
    $\thetab$ is a gain representing the emission variation over time.
    Except for the FBP reconstruction, only the product
    $\xb\circ\thetab$ has physical dimensions.  The spatial scales
    are given in solar radii and centered on the solar axis of
    rotation.  (a), (b) and (c) are slices of 3D cubes at the same
    $z=0.1\, R_\odot$.
    Emission densities (arbitrary units) are scaled in the color bars
    in the right-end side of (a), (b), (c).
  }
  \label{fig:SimResults}
\end{figure}

We now compare our results (Figure \ref{fig:SimResults}) with a
filtered back-projection (FBP) algorithm.  This method is explained by
\inlinecite{Natterer86} and \inlinecite{Kak87}.

By comparing the simulation and the reconstruction in Figure
\ref{fig:SimResults}, we can see the quality of the temporal evolution
estimation.  The shape of the intensity curves is well reproduced
except for the first plume in the first ten time steps where the
intensity is slightly underestimated. This corresponds to a period
when plume 1 is hidden behind plume 2. Thus, our algorithm attributes
part of the plume 1 intensity to plume 2.  Let us note that this kind
of ambiguity will not arise in the case of observations from multiple
points of view such as STEREO/EUVI observations.  The indeterminacy of
the problem is due to its bilinearity discussed in Section
\ref{subsection:TemporalEvolution}.  This allows the algorithm to
attribute larger values to the $\thetab$ parameters and to compensate
by decreasing the corresponding $\xb$.  This is not a drawback of the
method since it allows discontinuities between plumes and interplumes.
The only physical value of interest is the product $\xb\circ\gb$.

\begin{figure}
  \centering
  \begin{tabular}{ccc}
    \includegraphics[width=.29\linewidth]{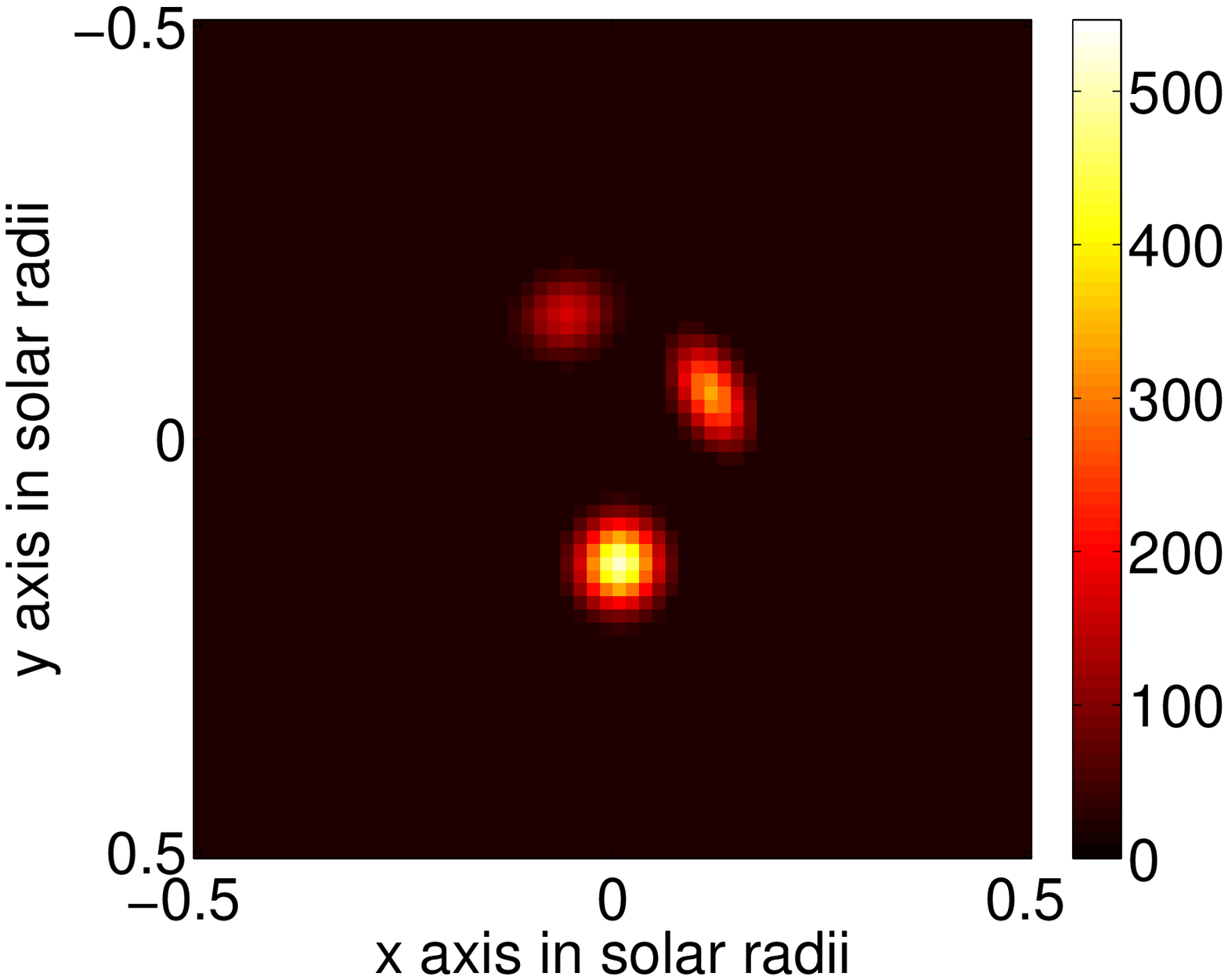}&
    \includegraphics[width=.29\linewidth]{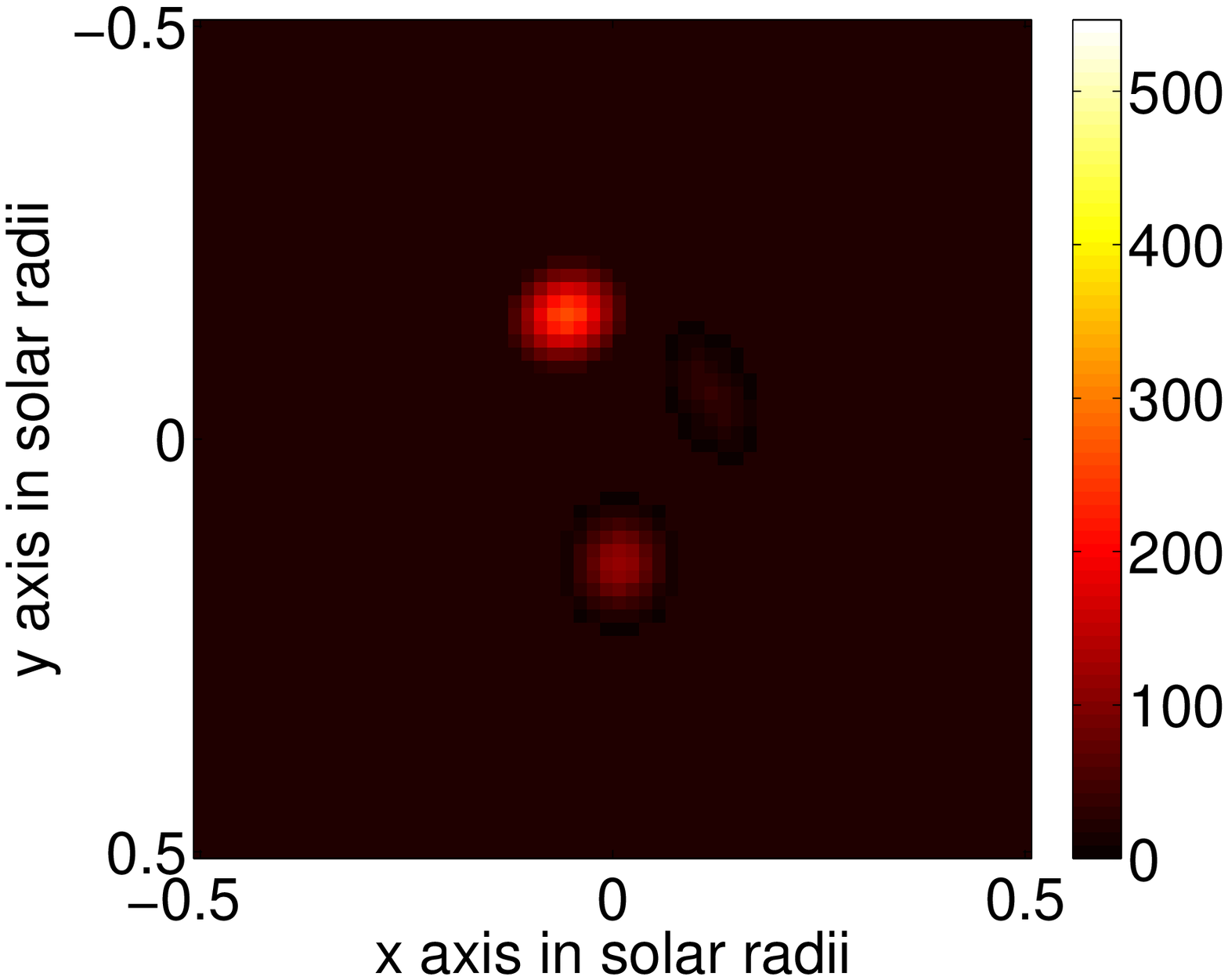}&
    \includegraphics[width=.29\linewidth]{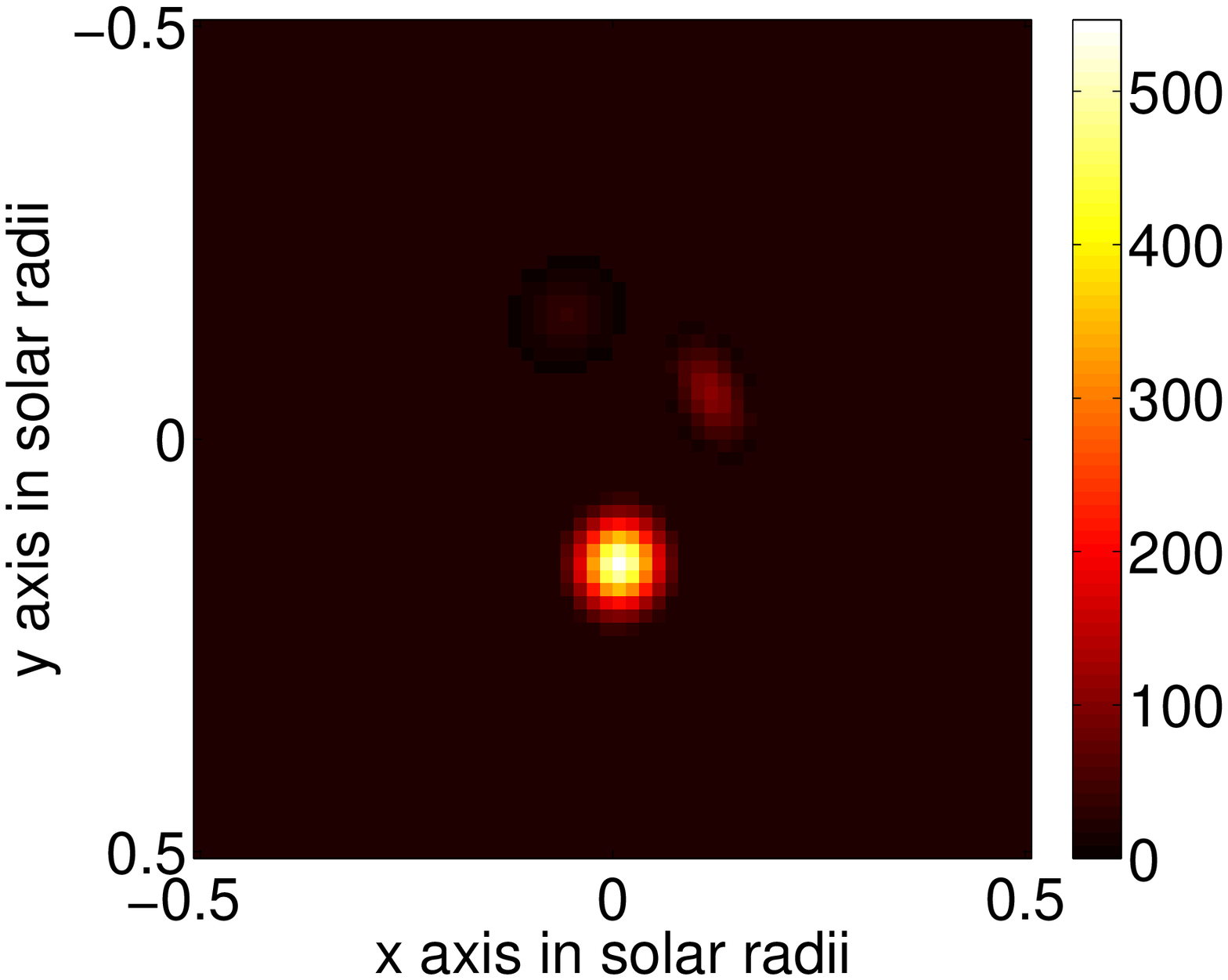}\\
    \footnotesize{(a) Simulation at $10 \Delta T$}&
    \footnotesize{(b) Simulation at $30 \Delta T$}&
    \footnotesize{(c) Simulation at $50 \Delta T$}\\
    
    \includegraphics[width=.29\linewidth]{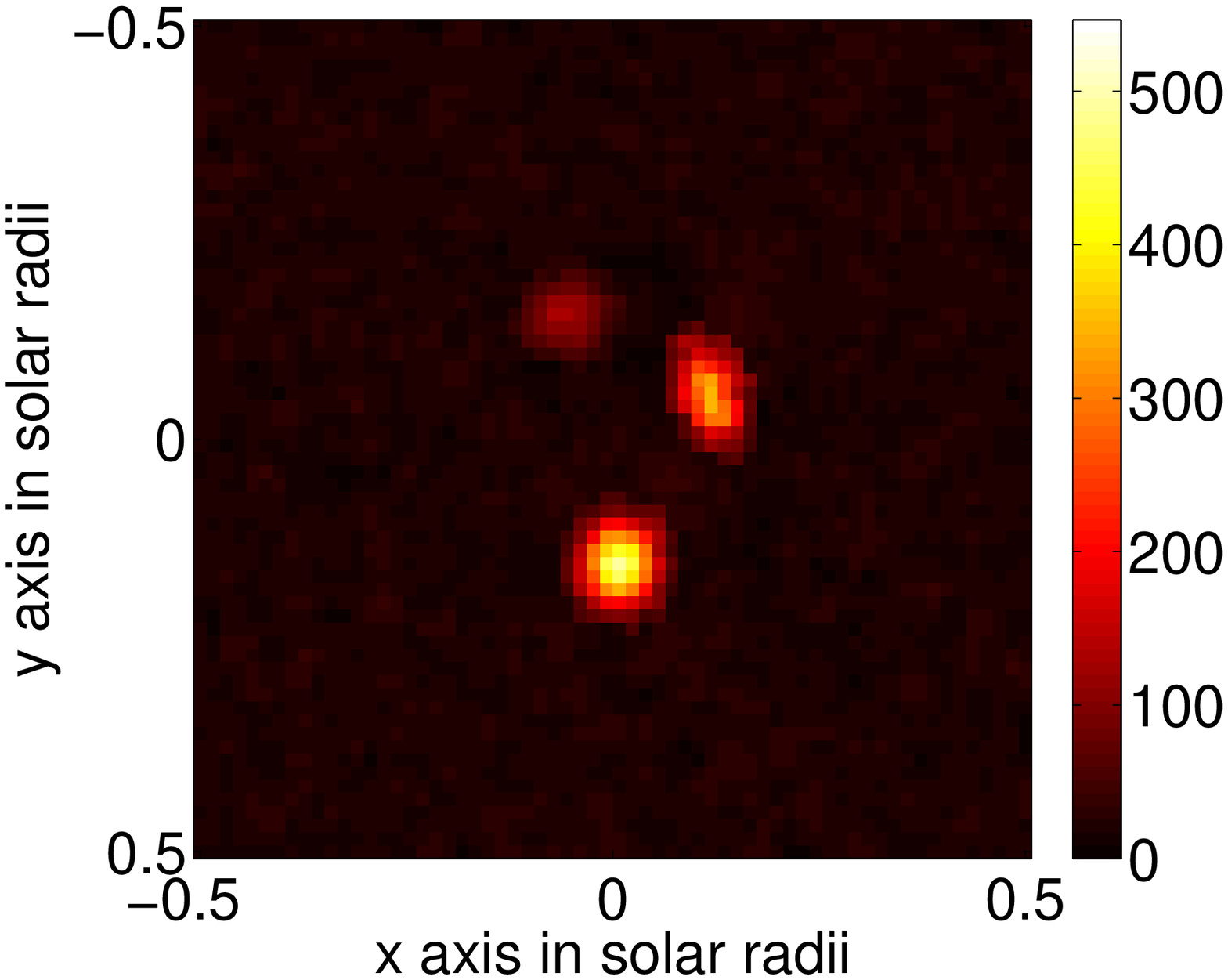}&
    \includegraphics[width=.29\linewidth]{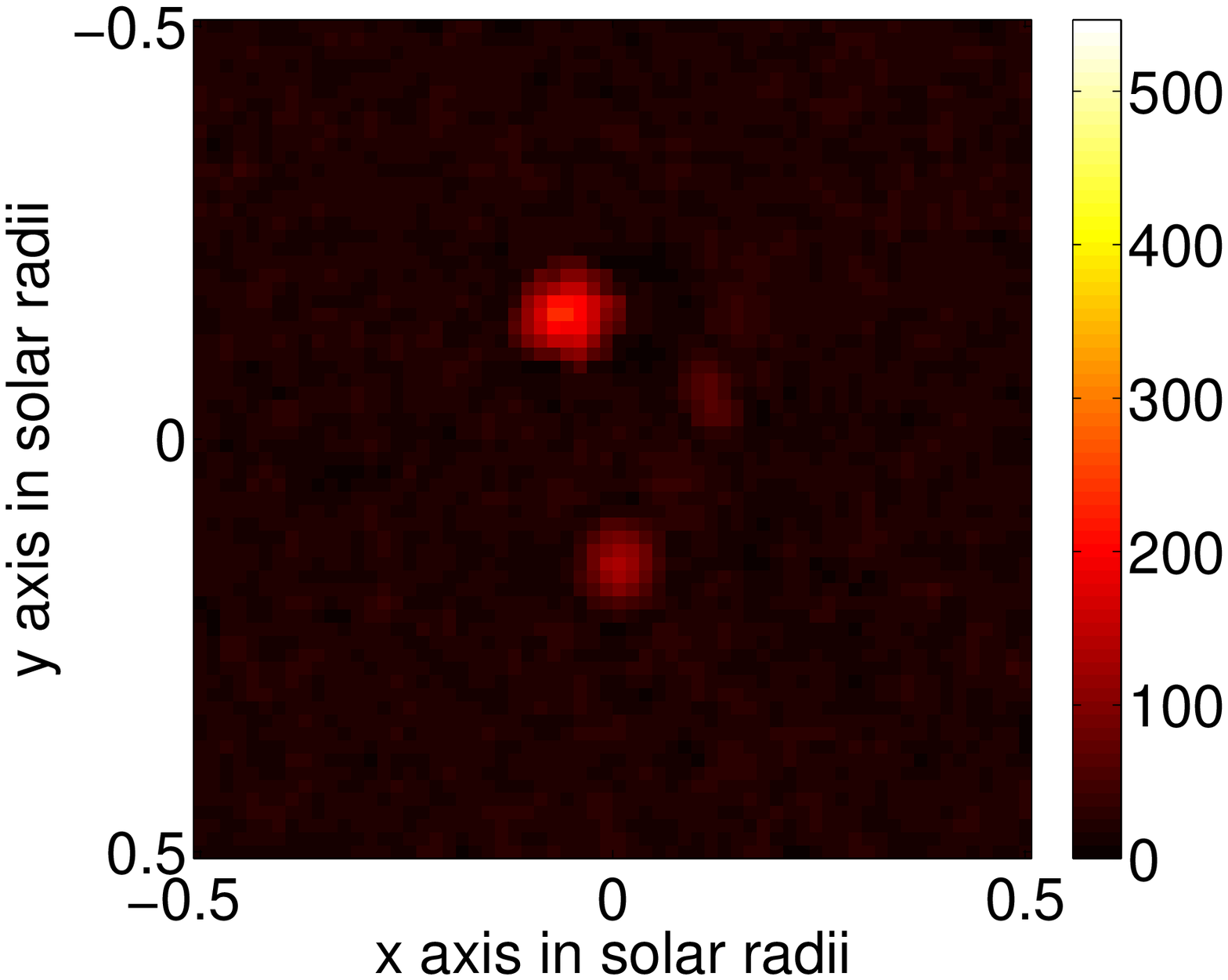}&
    \includegraphics[width=.29\linewidth]{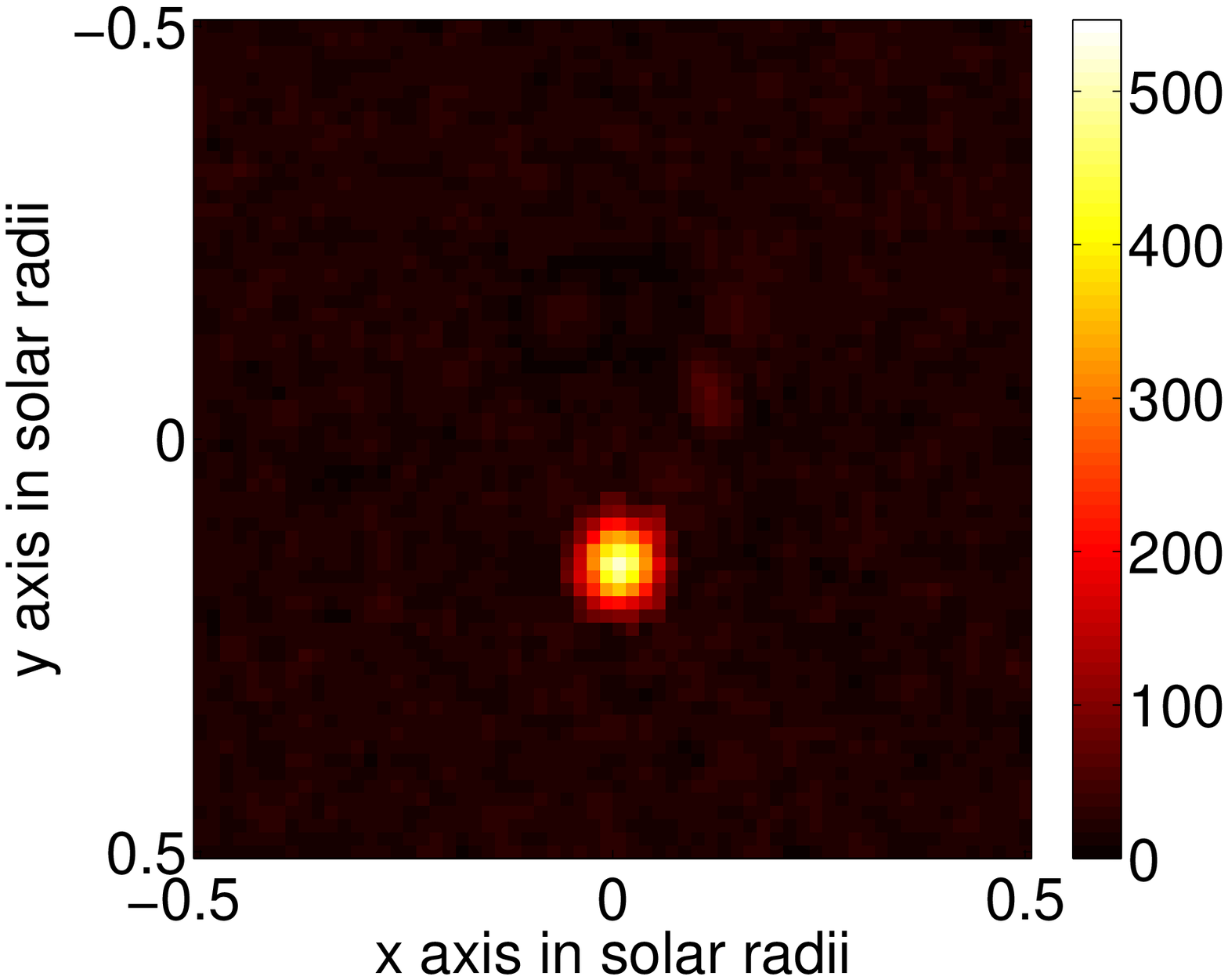}\\
    \footnotesize{(d) Result at $10 \Delta T$}&
    \footnotesize{(e) Result at $30 \Delta T$}&
    \footnotesize{(f) Result at $50 \Delta T$}\\
    
  \end{tabular}
  \caption
  { 
    Comparison of $\xb\circ\gb$ simulated and reconstructed at different times.
    $\Delta T$ is the time between two data images (5.6 hours).
    Distances are in solar radii. Values represent the volume emission density.
    All of this images are slices of 3D cubes at the same $z=0.1 R_\odot$.
  }
  \label{fig:compare}
\end{figure}

Figure \ref{fig:compare} shows the relative intensity of the plumes at
different times. One can compare with the reconstruction.  One way to
quantify the quality of the reconstruction is to compute the distance
(quadratic norm of the difference) between the real object and the
reconstructed one.  Since the FBP reconstruction does not actually
correspond to a reconstruction at one time, we evaluate the minimum of
the distances at each time.  We find it to be $3000$. This is to be
compared with a value of $700$ with our algorithm, which is much
better.

\subsection{Choice of Evolution Areas}
\label{subsection:areas}
\begin{figure}
  \centering
  \begin{tabular}{ccc}
    \includegraphics[width=.235\linewidth]{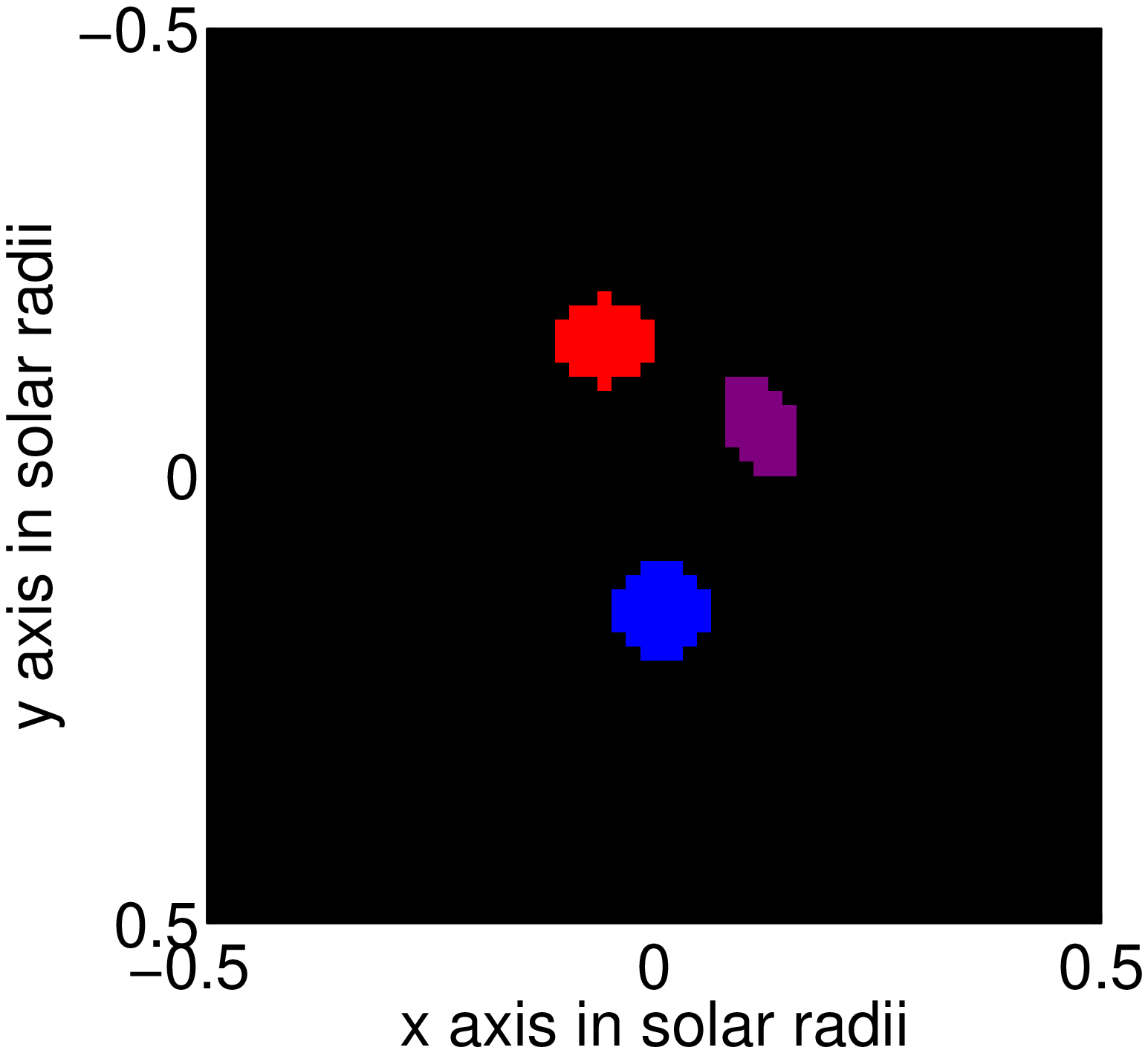}&
    \includegraphics[width=.27\linewidth]{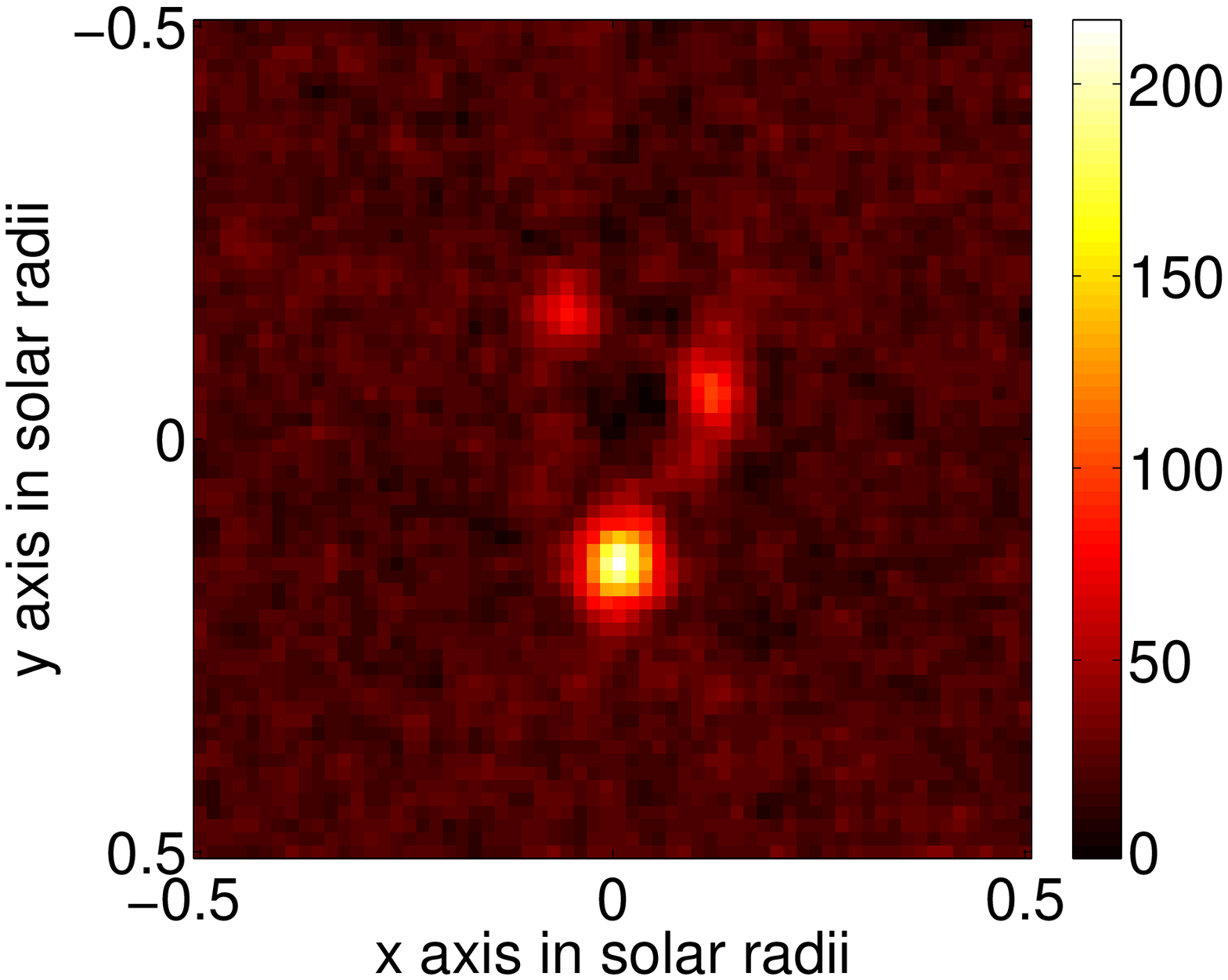}&
    \includegraphics[width=.27\linewidth]{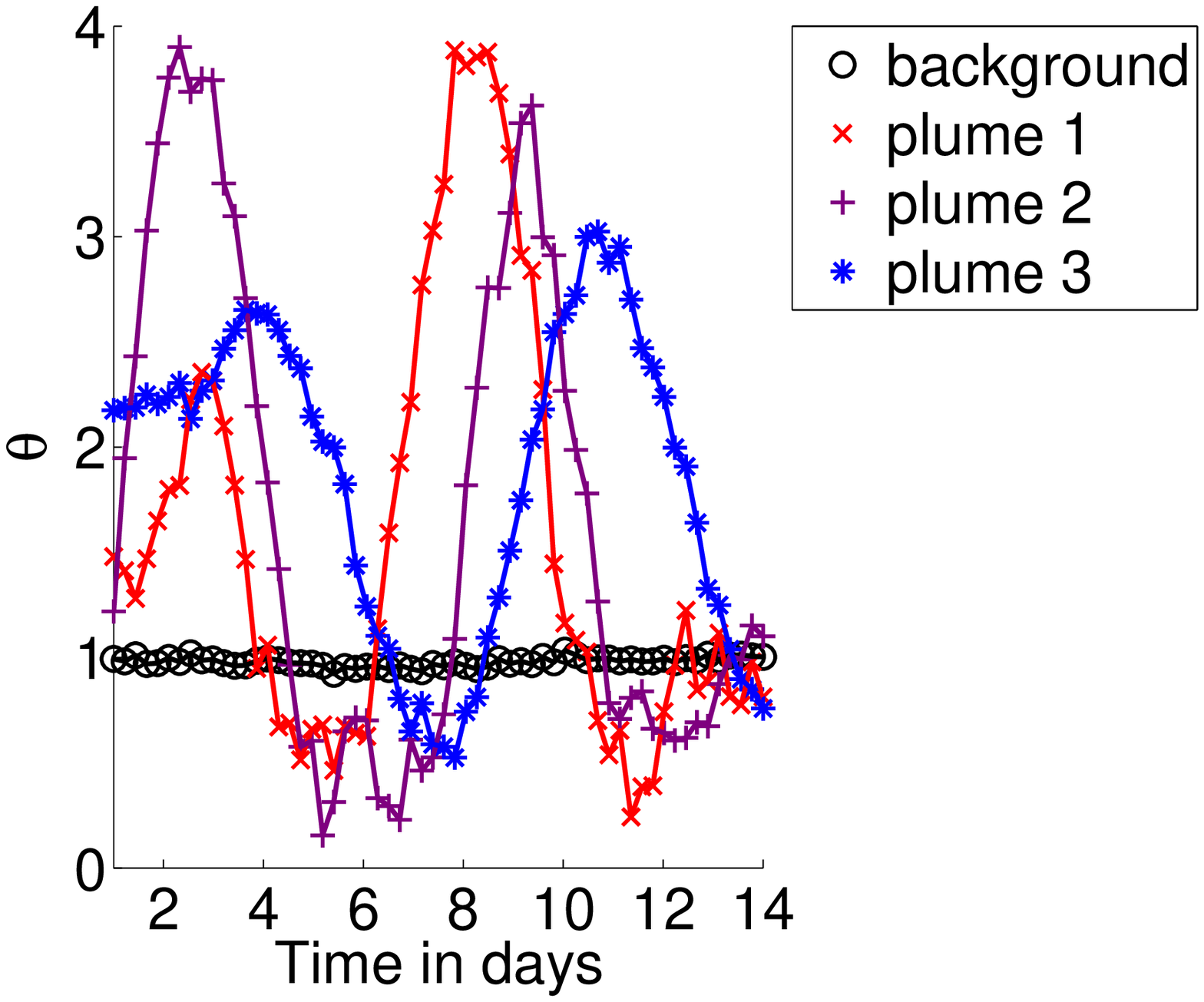}\\    
    \footnotesize{(a) New choice of areas}&
    \footnotesize{(b) $\xb$ with our algorithm}&
    \footnotesize{(c) $\thetab$ with our algorithm}\\
  \end{tabular}
  \caption
      {
        Reconstruction with smaller areas.
        To be compared with Figure \ref{fig:SimResults}.
        The new areas (a) do not correspond anymore to the ones used to
        generate the data.
        (b) is the emission map and (c) the temporal evolution estimated
        with our algorithm.
	(b) and (c) are slices of 3D cubes at the same $z=0.1\, R_\odot$.
        Emission densities (arbitrary units) are scaled in the color bars
        in the right-end side of (b).
      }
  \label{fig:SimFalseResults}
\end{figure}

One can think that the choice of the evolution areas is critical to the good
performance of our method. We show in this section that it is not
necessarily the case by performing a reconstruction based on simulations with
incorrect evolution areas. 
All parameters and data are exactly the same as in the
previous reconstruction. The only difference is in the choice of the areas, 
\textit{i.e.} the $\Lb$ matrix. These are now defined as shown in Figure 
\ref{fig:SimFalseResults}\textit{(a)}.

Although approximately 50 \% of the voxels are not associated with their 
correct area, we can observe that the algorithm still performs well.
The emission map of Figure \ref{fig:SimFalseResults}\textit{(b)} is still better
than the emission reconstructed by a FBP method. Plus, the estimation of the 
temporal evolution in Figure \ref{fig:SimFalseResults}\textit{(c)} corresponds to 
the true 
evolution \ref{fig:SimResults}\textit{(e)} even if less precisely than in Figure
\ref{fig:SimResults}\textit{(f)}.

\section{Reconstruction of SOHO/EIT Data}
\label{section:EITdata}

\subsection{Data Preprocessing}
We now perform reconstruction using SOHO/EIT data. We have to be
careful when applying our algorithm to real data. Some problems may
arise due to phenomena not taken into account in our model; \textit{e.g.}
cosmic rays, or missing data.

Some of these problems can be handled with simple preprocessing.  We
consider pixels hit by cosmic rays as missing data. They are detected
with a median filter.  These pixels and missing blocks are labeled as
missing data and the projector and the backprojector do not take them
into account (\textit{i.e.} the corresponding rows in the matrices are
removed).

\subsection{Results Analysis}

\begin{figure}
  \centering
  \begin{tabular}{cc}
    \includegraphics[width=.45\linewidth]{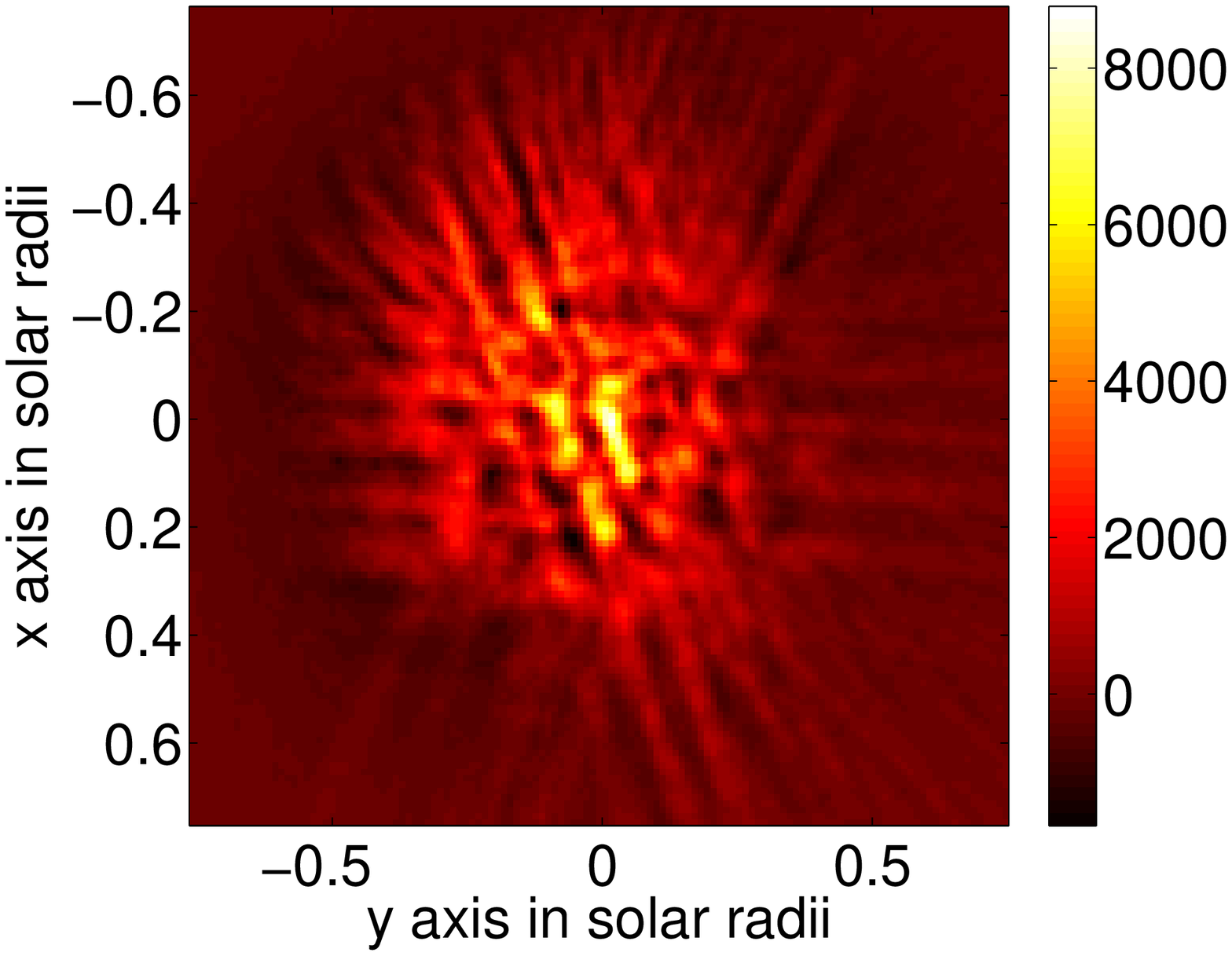}&
    \includegraphics[width=.37\linewidth]{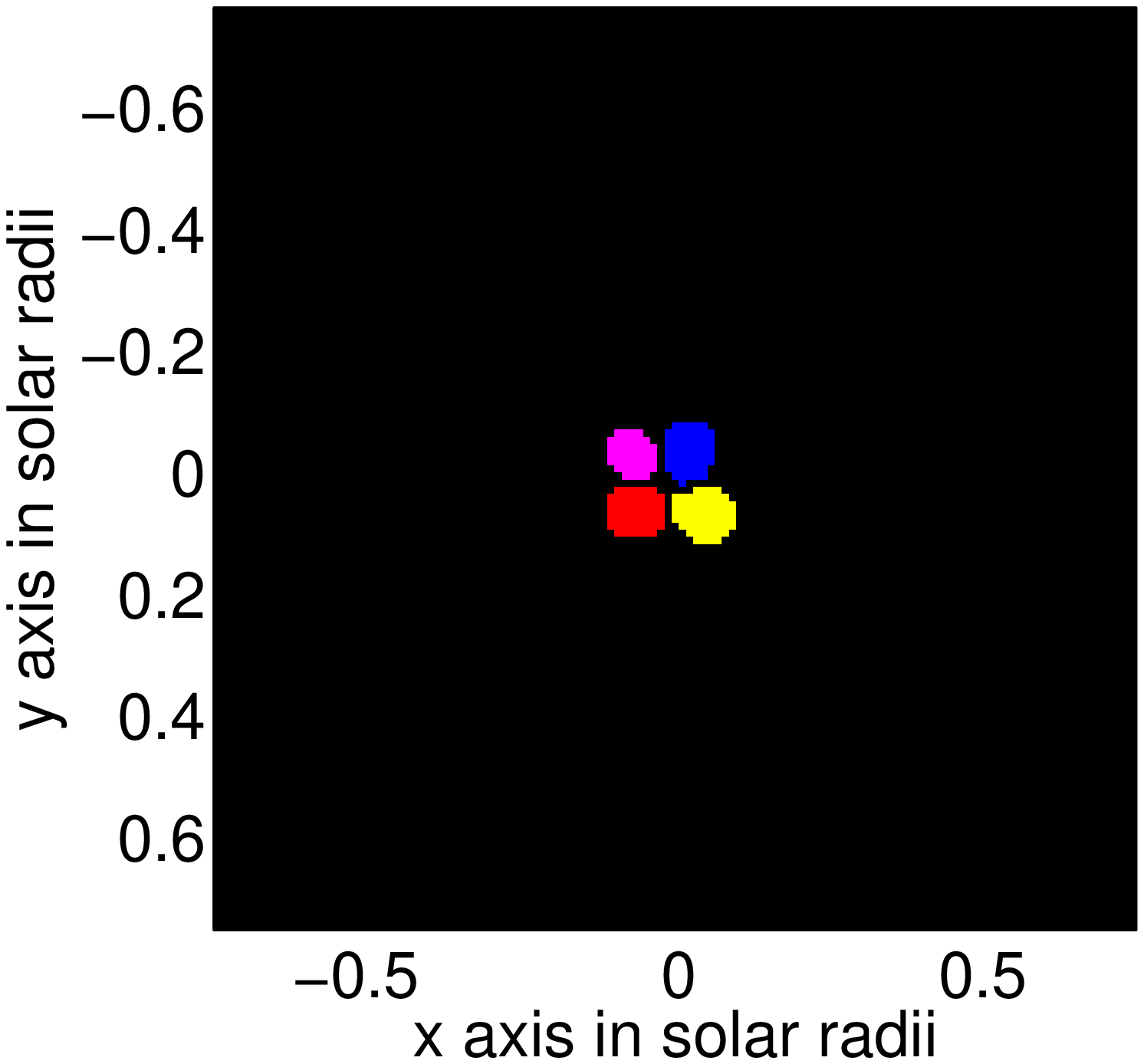}\\
    \footnotesize{(a) $\xb$ with a FBP algorithm}&
    \footnotesize{(b) Chosen temporal areas. Numbered}\\
    &\footnotesize{clockwise starting from lower-left}\\

    \includegraphics[width=.45\linewidth]{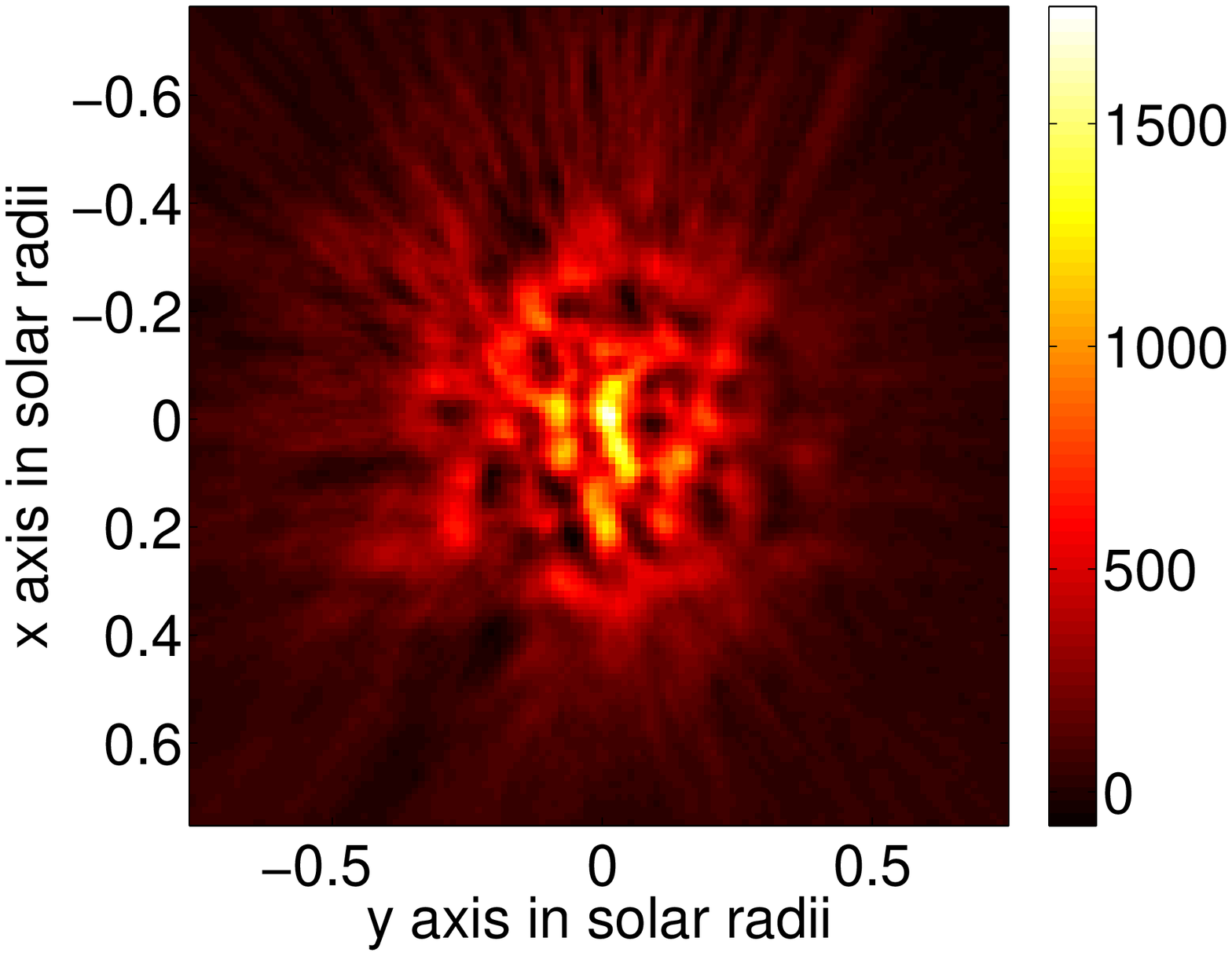}&
    \includegraphics[width=.45\linewidth]{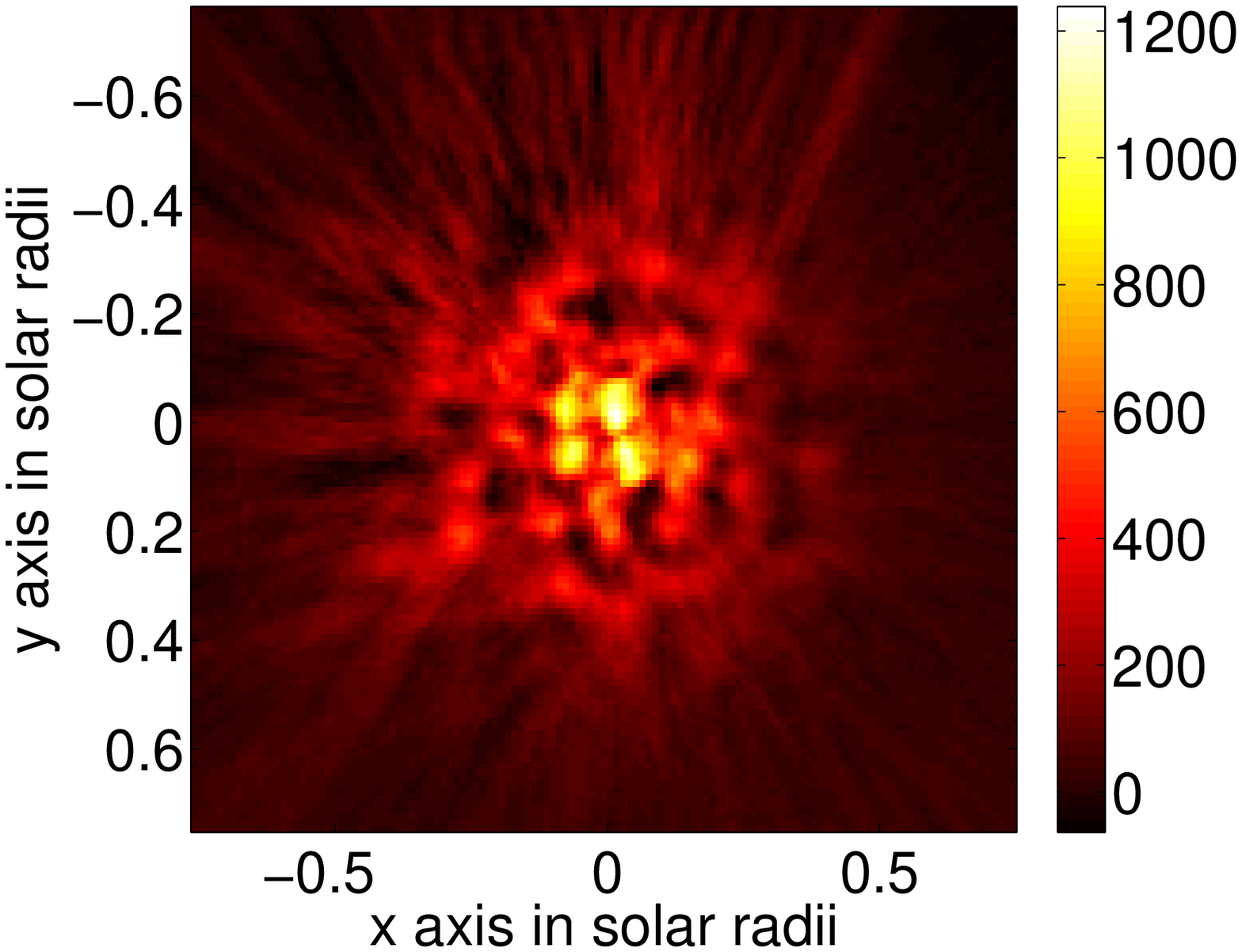}\\
    \footnotesize{(c) $\xb$ without temporal evolution}&
    \footnotesize{(d) $\xb$ with our algorithm}\\
    
  \end{tabular}
  \includegraphics[width=.9\linewidth,height=.45\linewidth]{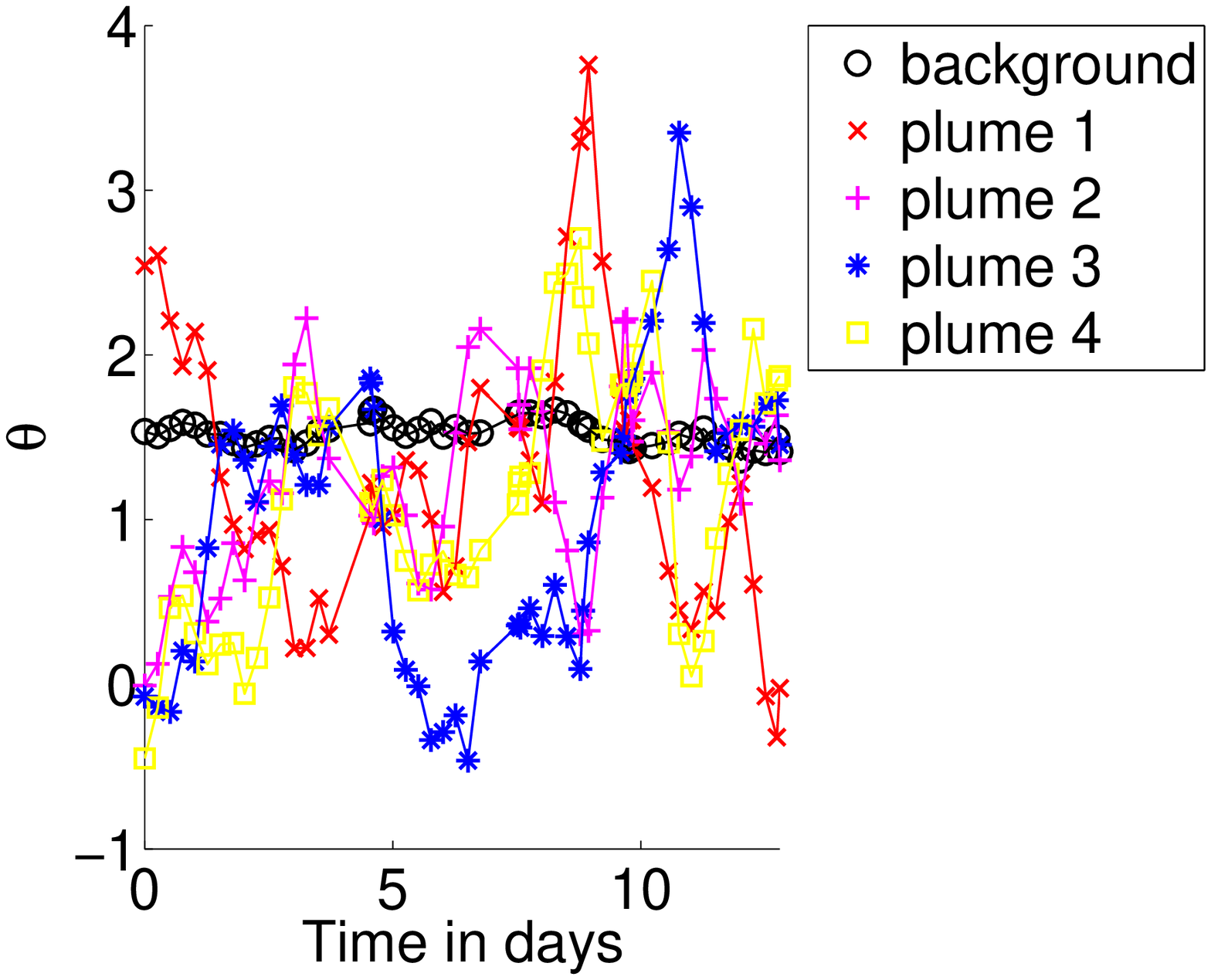}\\
  \footnotesize{(e) $\thetab_t$ with our algorithm.}
  \caption
  { A comparison of FBP (a), a gradient-like algorithm without
    temporal evolution (c), and our algorithm (d) with real EIT data.
    $\xb$ is the spatial distribution of the volume emission density
    integrated over EIT 17.1 nm passband.  The chosen areas are
    shown in (b).  $\thetab$ is a gain representing the emission
    variation during time (e).  The time scale is in days.  In the case
    of our algorithm, only the product $\xb\circ\thetab$ has
    physical meaning.  The spatial scales are given in solar radii
    and centered on the solar axis of rotation.  (a), (b), (c), and (d)
    are slices of 3D cubes at the same $z=1.3 R_\odot$.
    Emission densities (arbitrary units) are scaled in the color bars
    in the right-end side of (a), (c), (d).
  }
  \label{fig:SohoResults}
\end{figure}

\begin{figure}
  \centering
  \begin{tabular}{ccc}
    \includegraphics[width=.29\linewidth]{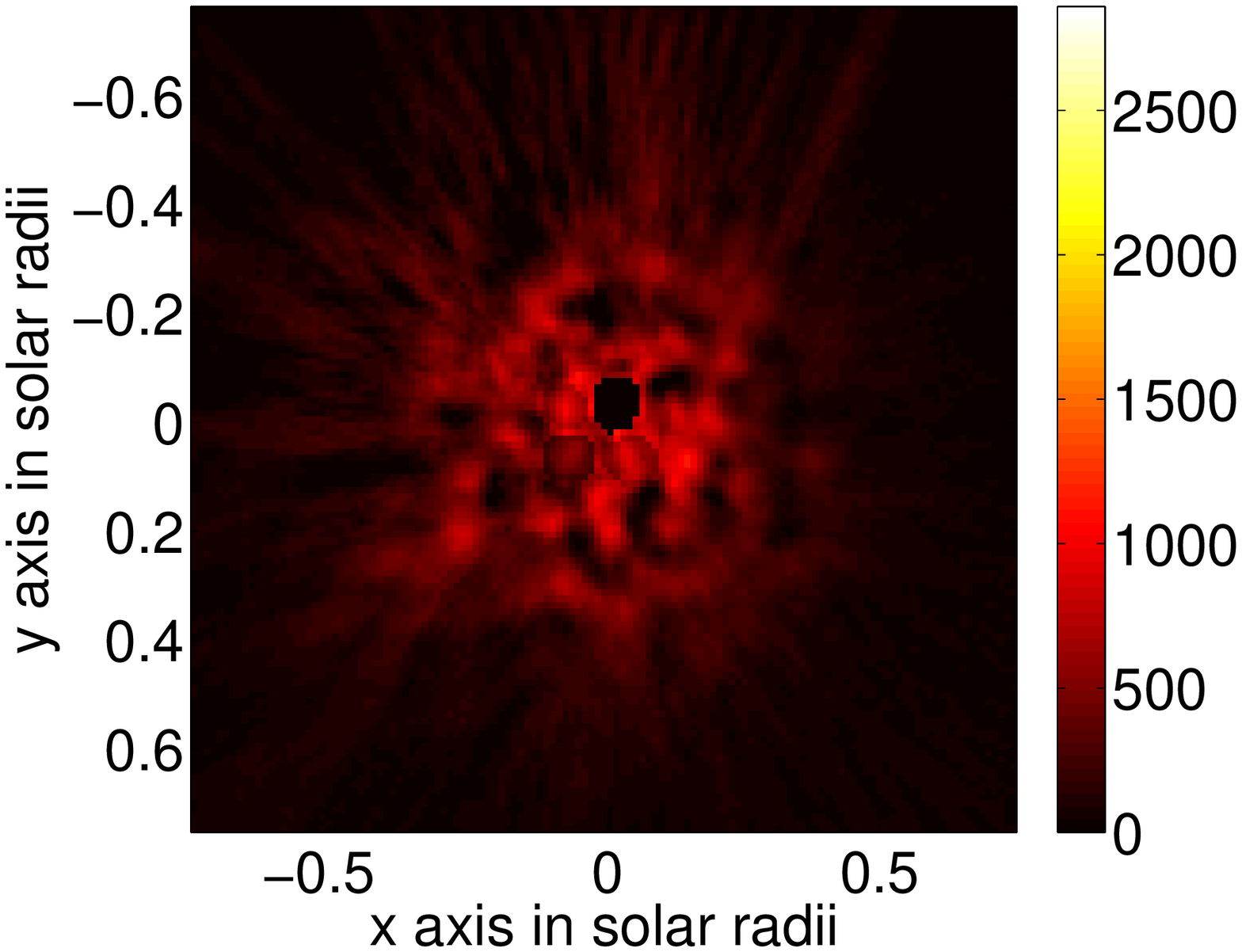}&
    \includegraphics[width=.29\linewidth]{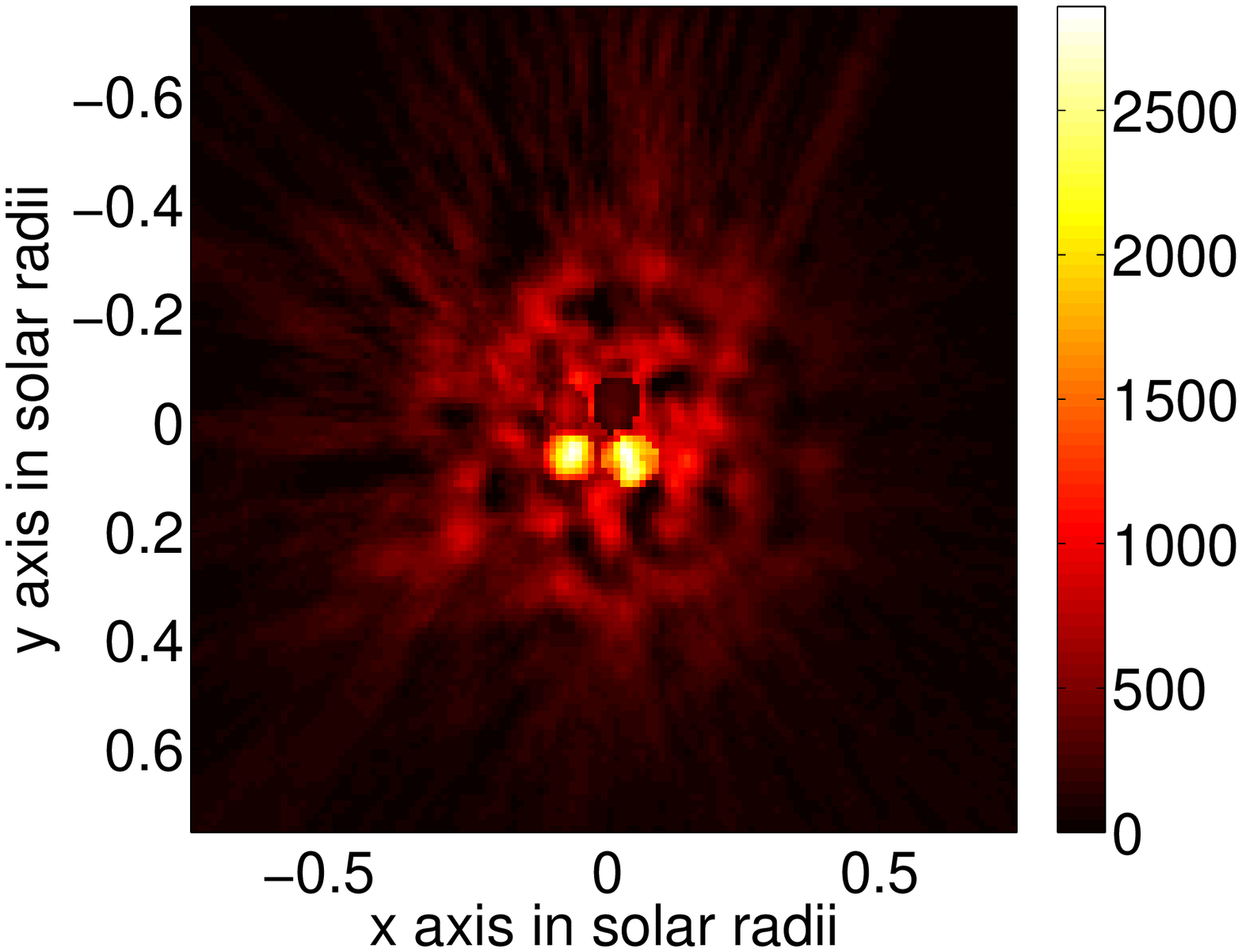}&  
    \includegraphics[width=.29\linewidth]{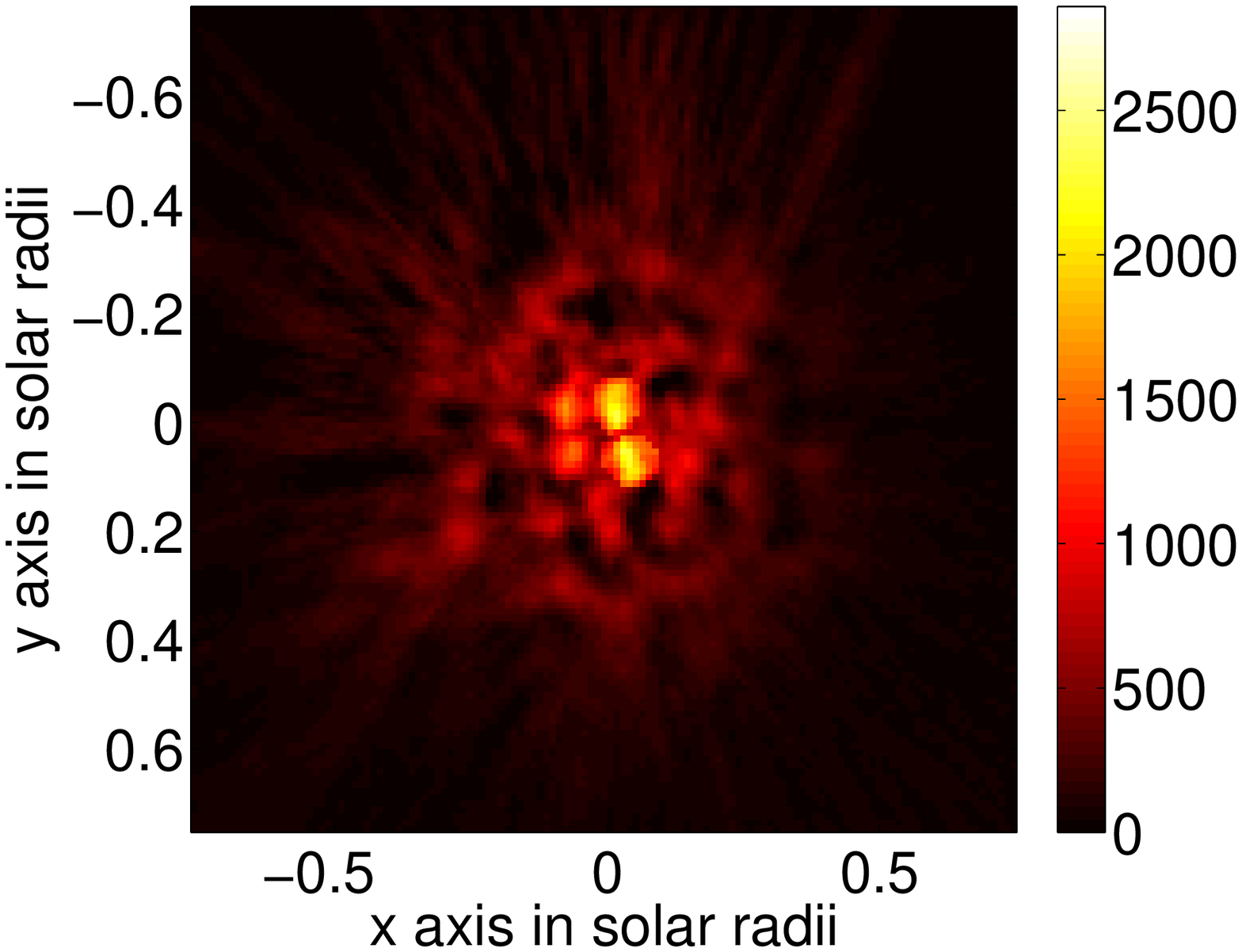}\\
    \footnotesize{(a) Reconstruction at $t_{25}$}&
    \footnotesize{(b) Reconstruction at $t_{35}$}&
    \footnotesize{(c) Reconstruction at $t_{45}$}\\
    \footnotesize{(6 days)}& 
    \footnotesize{(8.5 days)}&
    \footnotesize{(9.8 days)}\\
  \end{tabular}
  \caption
  { Reconstruction of $\xb\circ\gb$ at different times.  Distances
    are in solar radii.  Values represent the volume emission density
    integrated over the EIT 17.1 nm passband.  All of these images are
    slices of 3D cubes at the same $z=1.3\, R_\odot.$ }
  \label{fig:SohoCompare}
\end{figure}

In Figures \ref{fig:SohoResults} and \ref{fig:SohoCompare}, we present
results from 17.1 nm EIT data between 1 and 14 November
1996.  This period corresponds to the minimum of solar activity when
one can expect to have less temporal evolution. 17.1 nm is the
wavelength where the contrast of the plumes is the strongest.  Some
images are removed resulting in a sequence of 57 irregularly-spaced
projections for a total coverage of 191$^{\circ}$.  We assume that we
know the position of four evolving plumes as shown on Figure
\ref{fig:SohoResults}\textit{(b)}.  For each reconstructed image, we present
subareas of the reconstructed cube of size 64$\times$64 centered on
the axis of rotation.  We assume the rotation speed to be the rigid
body Carrington rotation.  All of the parameters given in Table
\ref{tab:SohoGeometricParameters} and \ref{tab:SohoOtherParameters}
are shared by the different algorithms provided they are required by
the method.  The computation of this reconstruction on a
Intel(R) Pentium(R) 4 CPU 3.00 GHz was 13.5 hours long.

\begin{table}
  \begin{tabular}{*{4}{c}}
    \hline
    cube size & cube number & pixel & projection \\
    (solar radii) & of voxels & size (radians)  & number of pixels\\
    \hline
    $3\times3\times0.15$ & $256\times256\times8$ & 
    $2.55\times 10^{-5} \times 2.55\times 10^{-5}$ & $512\times38$ \\
    \hline
  \end{tabular}
  \caption{EIT Data Reconstruction: Geometric Parameters}
  \label{tab:SohoGeometricParameters}
\end{table}

\begin{table}
  \begin{tabular}{*{4}{c}}
    \hline
    $\lambda$ & $\mu$ & $S_\xb$ & $S_G$\\
    \hline
    $2\times 10^{-2}$ & $1\times 10^4$ & 0.1 & 0.05\\
    \hline
  \end{tabular}
  \caption{EIT Data Reconstruction : Other Parameters}
  \label{tab:SohoOtherParameters}
\end{table}

Presence of negative values is the indication of a poor behavior of
the tomographic algorithm since it does not correspond to actual
physical values.  We can see in Figure \ref{fig:SohoResults} that our
reconstruction has many fewer negative values in the $\xb$ map than
the FBP reconstruction. In the FBP reconstruction cube, 50\% of the
voxels have negative values; in the gradient-like reconstruction
without temporal evolution 36\% of the voxels are negative while in
our reconstruction only 25 \% are negative.  This still seems like a
lot but most of these voxels are in the outer part of the
reconstructed cube. The average value of the negative voxels is much
smaller also. It is -120 for the FBP, -52 for the gradient-like method
without temporal evolution, and only -19 for our reconstruction with
temporal evolution. However, we notice that the gain coefficients
present a few slightly negative values.

In the reconstructions without temporal evolution, plumes three (upper
right) and four (lower right) correspond to a unique elongated
structure which we choose to divide. Note how our algorithm updated
the $\xb$ map reducing the emission values between these two plumes.
It shows that what was seen as a unique structure was an artifact
resulting from temporal evolution and it tends to validate the
usefulness of our model.  We note the disappearance of a plume
located around (-0.2, -0.15) solar radii on the FBP reconstruction. It
shows the utility of gradient-like methods to get rid of artifacts
due to the non-uniform distribution of images.  Another plume at (0.2,
0.2) solar radii has more intensity in the reconstruction without
temporal evolution than with our algorithm.  It illustrates how
temporal evolution can influence the spatial reconstruction.


\section{Discussion}
\label{section:discussion}
The major feature of our approach is the quality of our
reconstruction, which is much improved with respect to FBP reconstruction, as
demonstrated by the smaller number of negative values and the increased
closeness to the data.  Let us now discuss the various assumptions
that have been made through the different steps of the method.

The strongest assumption we made, in order to estimate the temporal
evolution of polar plumes, is the knowledge of the plume position.
Here, we choose to define the plumes as being the brightest points in
a reconstruction without temporal evolution.  The choice is not based
on any kind of automatic threshold. The areas are entirely hand-chosen
by looking at a reconstruction.  It is possible that these areas
do not correspond to the actual physical plumes, they could correspond
to areas presenting increased emission during half a rotation. Note
that this is biased in favor of plumes closer to the axis of rotation
since, along one slice of the reconstructed cartesian cube, their
altitude is lower and thus, their intensity is higher.  In order to
have constant altitude maps one would have to carry out the
computation on a spherical grid or to interpolate afterwards onto such
a grid.  For this reconstruction example we are aware that we did not
locate all of the plumes but only tried to find a few.  It would be
interesting to try to locate the plumes using other data or with a
method estimating their positions and shapes.

The method involves hyperparameters which we choose to set manually.
There are methods to estimate hyperparameters automatically such as
the L-curve method, the cross\--vali\-da\-tion method \cite{Golub79}
or the full-bayesian method \cite{Higdon97,Champagnat96a}.
We performed reconstructions using different hyperparameter values.
We then looked at the reconstruction to see if the smoothness seemed
exaggerated or if the noise were amplified in the results. This
allowed us to reduce the computational cost and does not really put
the validity of the method into question.

One possible issue with this algorithm is the non-convexity of our
criterion.  This can lead to the convergence to a local minimum that
does not correspond to the desired solution defined as the global
minimum of the criterion.  One way to test this would be to change the
initialization many times. 

We chose the speed of rotation of the poles to be the Carrington
rotation speed. But the speed of the polar structures has not been
measured precisely to our knowledge and could affect drastically the
reconstruction. This is an issue shared by all tomographic
reconstructions of the Sun.

In the current approach, we need to choose on our own the position of the
time-evolving areas which are assumed to be plumes. This
is done by assuming that more intense areas of a reconstruction
without temporal evolution correspond to plume positions.  A more
rigorous way would be to try to use other sources of information to
try to localize the plumes. Another, self-consistent way, would be to develop a
method that jointly estimates the position of the plumes in addition
to the emission ($\xb$) and the time evolution ($\thetab$).
We could try to use the results of \citeauthor{Yu02} (\citeyear{Yu02})
who propose an original approach in order to reconstruct a 
piece-wise homogeneous object while preserving edges.
The minimization is alternated between an intensity map and boundary
curves. The estimation of the boundary curves is made using level sets
technics (\cite{Yu02} and references therein).
It would also be possible to use a Gaussian mixture model \cite{snoussi07c}.

\section{Conclusion}
\label{section:conclusion}
We have described a method that takes into account the temporal
evolution of polar plumes for tomographic reconstruction near the
solar poles.  A simple reconstruction based on simulations
demonstrates the feasibility of the method and its efficiency in
estimating the temporal evolution assuming that parameters such as
plume position or rotation speed are known.  Finally we show that it
is possible to estimate the temporal evolution of the polar plumes
with real data.

In this study we limited ourselves to reconstruction of images at 17.1
nm but one can perform reconstructions at 19.5 nm and 28.4 nm as well.
It would allow us to estimate the temperatures of the electrons as in
\citeauthor{Frazin05} (\citeyear{Frazin05}) or 
Barbey \etal (\citeyear{Barbey06}).

\begin{acks}
Nicolas Barbey acknowledges the support of the 
Centre National \\ d'{\'E}tudes Spatiales
and the \CLS.
The authors thank the referee for their useful suggestions for the
article.
\end{acks}

\appendix
\section{Pseudo-Inverse Minimization}
\label{app:inverse}

We want to minimize:
\begin{equation}
  J = \|\yb - \Ub_{\xb^n}\thetab \|^2
  + \lambda\|\Db_r \xb^n \|^2 + \mu\|\Db _t\thetab\|^2
\end{equation}
The second term does not depend on $\thetab$.  Due to the strict
convexity of the criterion, the solution is a zero of the gradient.
Since the criterion is quadratic, one can explicitly determine the
solution:
\begin{equation}
  \label{eq:gradient}
  \left. \grad_{\thetab} J\right|_{\thetab = \thetab^{n+1}} = 
  2\Ub_{\xb^n}^T \left ( \Ub_{\xb^n} \thetab^{n+1} - \yb \right)
  + 2 \mu \Db_t^T\Db_t \thetab^{n+1}
  = \mathbf{0}
\end{equation}
from which we conclude:
\begin{equation}
  \label{eq:thetamin}
  \thetab^{n+1} = \left[ \Ub_{\xb^n}^T \Ub_{\xb^n} +
    \mu \Db_t^T \Db_t \right]^{-1}\Ub_{\xb^n}^T \yb
\end{equation}

\section{Gradient-like Method}
\label{app:gradient}
In this method we try to find an approximation of the minimum by
decreasing the criterion iteratively.  The problem is divided in two
subproblems: searching for the direction and searching for the step of
the descent. In gradient-like methods, the convergence is generally
guaranteed ultimately to a local minimum.  But since the criterion is
convex, the minimum is global.  To iterate, we start at an arbitrary
point $(\xb^0)$ and go along a direction related to the gradient. The
gradient at the $p^{th}$ step is:
\begin{equation}
  \label{eq:gradp}
  \left. \grad_{\xb} J \right|_{\xb = \xb^p} = 
  2\Vb_{\thetab^{n+1}}^T \left ( \Vb_{\thetab^{n+1}} \xb^p - \yb \right) + 
  2 \lambda \Db_r^T\Db_r \xb^p
\end{equation}
Once the direction is chosen, searching for the optimum step is a
linear minimization problem of one variable:
\begin{equation}
  a\OPT^{p+1} = \argmin_{a} J(\xb^p + a \db^{p+1})
\end{equation}
which is solved by:
\begin{equation}
  \label{eq:step}
  a\OPT^{p+1} = -\frac{1}{2}\frac{\db^{p+1} \left. \grad_{\xb} J \right|_{\xb = \xb^p}}
  {\|\Vb_{\thetab^{n+1}}\db^{p+1}\|^2 + \lambda\|\Db_r\db^{p+1}\|^2}
\end{equation}
We can write the iteration:
\begin{equation}
  \label{eq:update}
  \xb^{p+1} = \xb^p + a\OPT^{p+1} \db^{p+1}
\end{equation}

\bibliographystyle{spr-mp-sola}
\bibliography{barbey_SPfullpaper}
\end{article}
\end{document}